\documentclass[twocolumn]{aastex701}

\usepackage{amsmath}

\turnoffeditone
\begin{document}

\title{A self-consistent solar coronal heating model by Alfv{\'e}nic waves}

\author[orcid=0000-0003-1134-2770, gname=Hidetaka, sname=Kuniyoshi]{Hidetaka Kuniyoshi}
\affiliation{School of Engineering, Physics and Mathematics, Northumbria University, UK}
\email[show]{hidetaka.kuniyoshi@northumbria.ac.uk}  

\author[orcid=0000-0001-5678-9002,gname=Richard, sname=Morton]{Richard J. Morton} 
\affiliation{School of Engineering, Physics and Mathematics, Northumbria University, UK}
\email{richard.morton@northumbria.ac.uk}

\begin{abstract}

Alfv{\'e}nic waves are prevalent throughout the solar atmosphere and are believed to play an essential role in coronal heating, classified as alternating current (AC) heating in contrast to direct current (DC) heating associated with quasi-static magnetic field line braiding.
The relative importance of AC versus DC heating depends on the details of the photospheric driver and on the configuration of the magnetic field. 
\added{Moreover, even if AC heating prevails, several wave dissipation mechanisms have been proposed, and which of them dominates remains unclear, as its efficiency depends on plasma compressibility and density inhomogeneity.}
We address these issues by performing three-dimensional radiative magnetohydrodynamic (MHD) simulations of a coronal loop spanning from the upper convection zone to the corona, which self-consistently capture many relevant physical processes.
We find that the corona is predominantly heated by AC heating, with Alfv{\'e}n wave turbulence providing the primary contribution, accounting for at least $80\%$ of the entire coronal heating \added{in the present simulation.}
Our results strongly support the use of Alfv{\'e}n wave turbulence--based models employed in space weather and stellar activity research, such as the Alfv{\'e}n Wave Solar Model (AWSoM) and the Magnetohydrodynamic Algorithm outside a Sphere (MAS).

\end{abstract}


\section{Introduction}\label{sec:introduction}

The solar corona is orders of magnitude hotter than the underlying photosphere, a phenomenon that remains poorly understood. 
Through interactions between magnetic fields and convective plasma motions, energy is transferred through the chromosphere into the corona \citep{Priest_2014_textbook}.
The heating mechanisms are broadly classified into alternating current (AC) and direct current (DC), depending on whether the timescale of magneto-convective driving is shorter or longer than the Alfv{\'e}nic wave crossing time of a coronal loop \citep{Aschwanden_2004_textbook, Klimchuk_2006_SoPh}. 
In the AC regime, energy is carried by Alfv{\'e}nic waves \citep{Alfven_1947_MNRAS, Uchida_1974_SoPh}; in the DC regime, it accumulates through gradual braiding of magnetic field lines \citep{Parker_1972_ApJ, Parker_1983_ApJ}. 
Since observed Alfv{\'e}nic wave periods span from $10\ \rm s$ to $10^4\ \rm s$ \citep[e.g.,][]{Tomczyk_2009_ApJ, Morton_2025a_ApJ} and thus can exceed the loop crossing time, the AC/DC boundary is not sharp in practice. 
Throughout this paper, we use ``AC heating'' to refer specifically to Alfv{\'e}nic wave heating.

Given that Alfv{\'e}nic waves have been ubiquitously observed throughout the corona over the past two decades \citep[e.g.,][]{Aschwanden_1999_ApJ, Nakariakov_1999_Sci, Tomczyk_2007_Sci, Morton_2025_nata}, the potential contribution of the AC heating has been extensively investigated \citep[see the reviews by, e.g.,][]{VanDoorsselaere_2020_SSRv, Morton_2023_RvMPP}.
One of the essential questions related to the AC heating is: how is Alfv{\'e}nic wave energy ultimately dissipated into heat? 
A number of dissipation mechanisms have been proposed, yet no clear consensus has been reached.

\medskip

% The Alfv{\'e}n wave turbulence model \citep{Hossain_1995_PhFl, Matthaeus_1999_ApJ} is perhaps the most widely applied, used to model solar coronal heating \citep{Nigro_2008_ApJ, vanBallegooijen_2011_ApJ, Verdini_2012_AA}, space weather 
% \citep{vanDerHolst_2014_ApJ, Mikic_2018_NatAs}, and stellar activity \citep{AlvaradoGomez_2016_AA, Shoda_2024_AA, Chen_2025_ApJ}. 

The Alfvén wave turbulence model \citep{Hossain_1995_PhFl} is perhaps the most widely applied, used to model the heating of coronal loops \citep[e.g.,][]{Buchlin_2007_ApJ, Nigro_2008_ApJ, vanBallegooijen_2011_ApJ, Verdini_2012_AA, Asgari_2012} and the solar wind plasma \citep{Matthaeus_1999_ApJ, Cranmer_2007_ApJS, Zank_2017, Squire_2022}.
Moreover, the model is widely applied for space weather 
\citep{vanDerHolst_2014_ApJ, Mikic_2018_NatAs}, and stellar magnetic activity \citep{AlvaradoGomez_2016_AA, Shoda_2024_AA, Chen_2025_ApJ}.
Alfv{\'e}n wave turbulence is mediated by nonlinear interactions of 
counter-propagating Alfv{\'e}nic waves \citep{Iroshnikov_1964_SvA, Kraichnan_1965_PhFl}, leading to an anisotropic energy cascade primarily in the perpendicular direction to the background magnetic field \citep{Goldreich_1995_ApJ, Boldyrev_2006}; see also \citet{Schekochihin_2022} for a comprehensive review.

\medskip

While the Alfv{\'e}n wave turbulence model generally neglects compressibility, Alfv{\'e}nic waves themselves can drive shock heating. In the chromosphere and transition region, strong Alfv{\'e}n speed gradients cause upward-propagating waves to grow in amplitude, undergo nonlinear mode conversion into compressive modes, and subsequently steepen into shocks, potentially contributing to coronal heating \citep{Hollweg_1982_SoPh, Moriyasu_2004_ApJ, Suzuki_2005_ApJ, 
Antolin_2010_ApJ, Arber_2016_ApJ}.

\medskip

The Alfv{\'e}n wave turbulence model also neglects density 
inhomogeneities especially perpendicular to the background 
magnetic field, yet observations indicate non-negligible 
perpendicular density contrasts ranging from $1.1$ to $10$ 
\citep{Aschwanden_2003_ApJ, AsensioRamos_2013_AA, 
Verwichte_2013_AA, Morton_2021_ApJ}. 
Here, the 
density contrast is defined as $\zeta = \rho_i / \rho_e$, 
where $\rho_i$ and $\rho_e$ are the 
plasma densities inside and outside the coronal loop, 
respectively.
Perpendicular density inhomogeneity enables several additional damping/dissipation mechanisms. 
In the phase mixing model \citep{Heyvaerts_1983_AA}, Alfv{\'e}nic wave fronts on neighboring field lines are progressively distorted at Alfv{\'e}n speed gradients, generating increasingly small spatial scales. 
Similarly, resonant absorption \citep{Ionson_1978_ApJ, Terradas_2010_AA} converts large-scale coherent Alfv{\'e}nic motions into localized oscillations near the resonant layer.
The uniturbulence model \citep{Magyar_2017_NatSR, Magyar_2019b_ApJ} demonstrates that perpendicular inhomogeneity enables a self-cascade of unidirectional Alfv{\'e}nic waves, removing the requirement for counter-propagating waves. Finally, the Kelvin--Helmholtz instability \citep{Terradas_2008_ApJ, 
Magyar_2016_AA, Antolin_2016_ApJ, Guo_2019_ApJ} can develop at density interfaces under velocity shear, generating small-scale dissipative structures. The efficiency of all these mechanisms depends on the magnitude of the density contrast.

\medskip

Despite continued interest in Alfv{\'e}nic wave heating, which dissipation mechanism dominates in the real corona remains unclear. Most previous analytical studies have assumed idealized conditions in which only a single mechanism operates \citep[see the review by][]{VanDoorsselaere_2020_SSRv}, and many numerical models employed simplified photospheric drivers 
\citep[e.g.,][]{Matsumoto_2018_MNRAS, Howson_2022_AA} or treated the density contrast as a free parameter \citep[e.g.,][]{Karampelas_2017_AA, Guo_2019_ApJ}. These approaches clarify individual mechanisms but cannot assess their relative 
efficiency simultaneously.

Magneto-convection simulations are three-dimensional radiative compressible magnetohydrodynamic (MHD) models spanning the upper convection zone to the corona \citep[e.g.,][]{Hansteen_2015_ApJ, Rempel_2017_ApJ, Chen_2022_ApJ} that self-consistently capture Alfv{\'e}nic wave generation at the surface, propagation through the chromosphere, and dissipation in the corona.
In addition, chromospheric evaporation and jets occur spontaneously \citep{Iijima_2017_ApJ}, allowing perpendicular density inhomogeneities in the corona to develop without ad-hoc assumptions \citep{Malanushenko_2022_ApJ, Kohutova_2024_AA}. These properties make magneto-convection simulations ideally suited to assess the relative efficiency of multiple Alfv{\'e}nic wave dissipation mechanisms simultaneously.

\medskip

In this paper, we perform a magneto-convection simulation 
representative of a quiet-Sun coronal loop rooted in network magnetic field regions, aiming to determine whether AC 
heating by Alfv{\'e}nic waves plays a significant role compared with DC 
heating, and if so, to identify which Alfv{\'e}nic wave dissipation mechanism is most efficient 
under realistic coronal density inhomogeneities.

\begin{figure*}[t!]
  \centering
  \includegraphics[width=15cm]{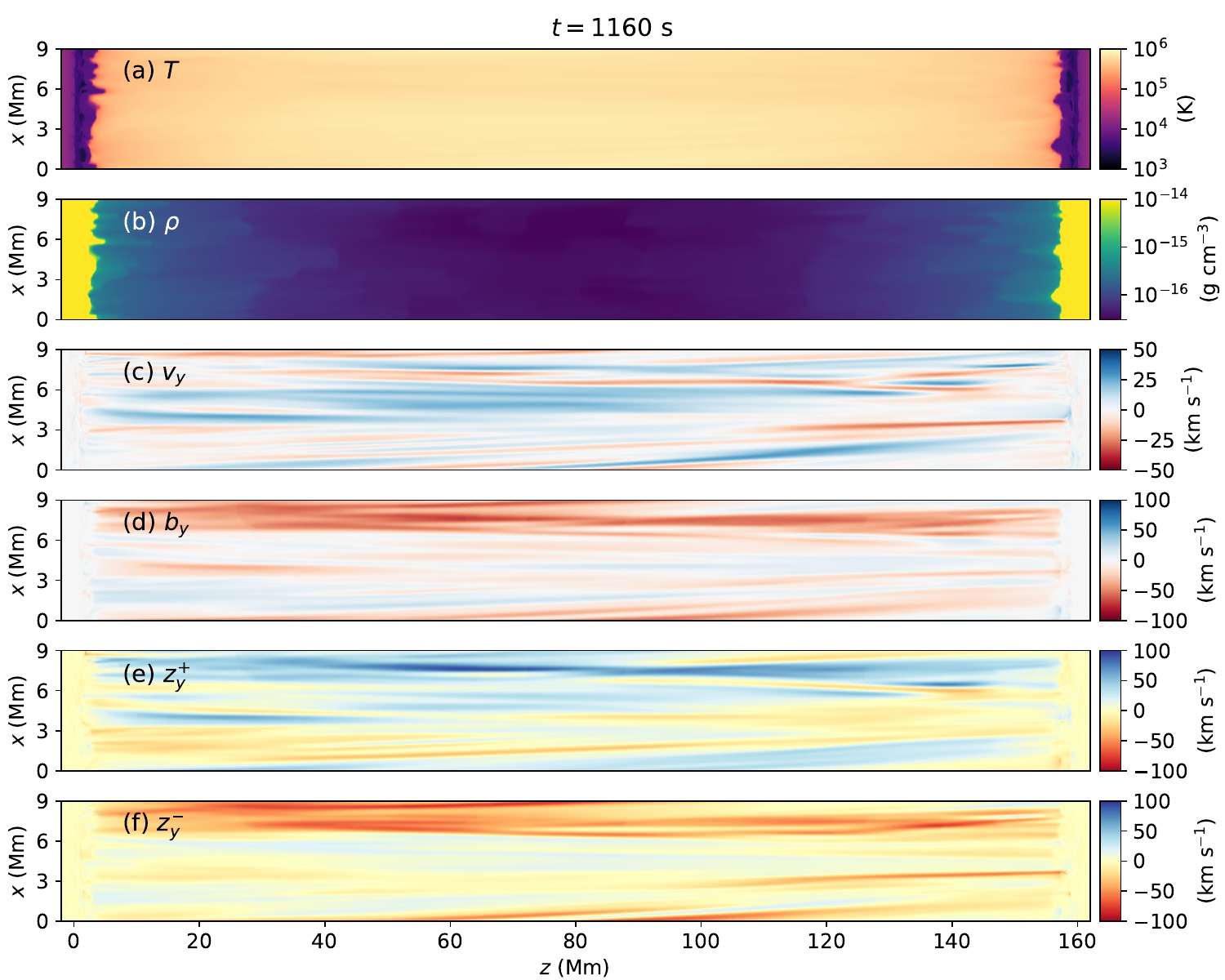} % 幅を10cmに指定
  \caption{
  Spatial distributions of physical quantities in the $xz$-plane at $t = 1160\ \rm s$, illustrating the overall structure of the simulated atmosphere. (a) Temperature $T$, (b) mass density $\rho$, (c) $y$-directional velocity $v_y$, (d) normalized magnetic field in velocity units $b_y=B_y/\sqrt{4\pi\rho}$, and (e)--(f) Els\"{a}sser variables $z^{\pm}_{y}$. The aspect ratio is not to scale to improve visibility. The associated animation shows the temporal evolution over a period from $t=0 \ \rm s$ to $t=5400 \ \rm s$.
  }
  \label{fig:snapshots}
\end{figure*}

\begin{figure*}[t!]
  \centering
  \includegraphics[width=14cm]{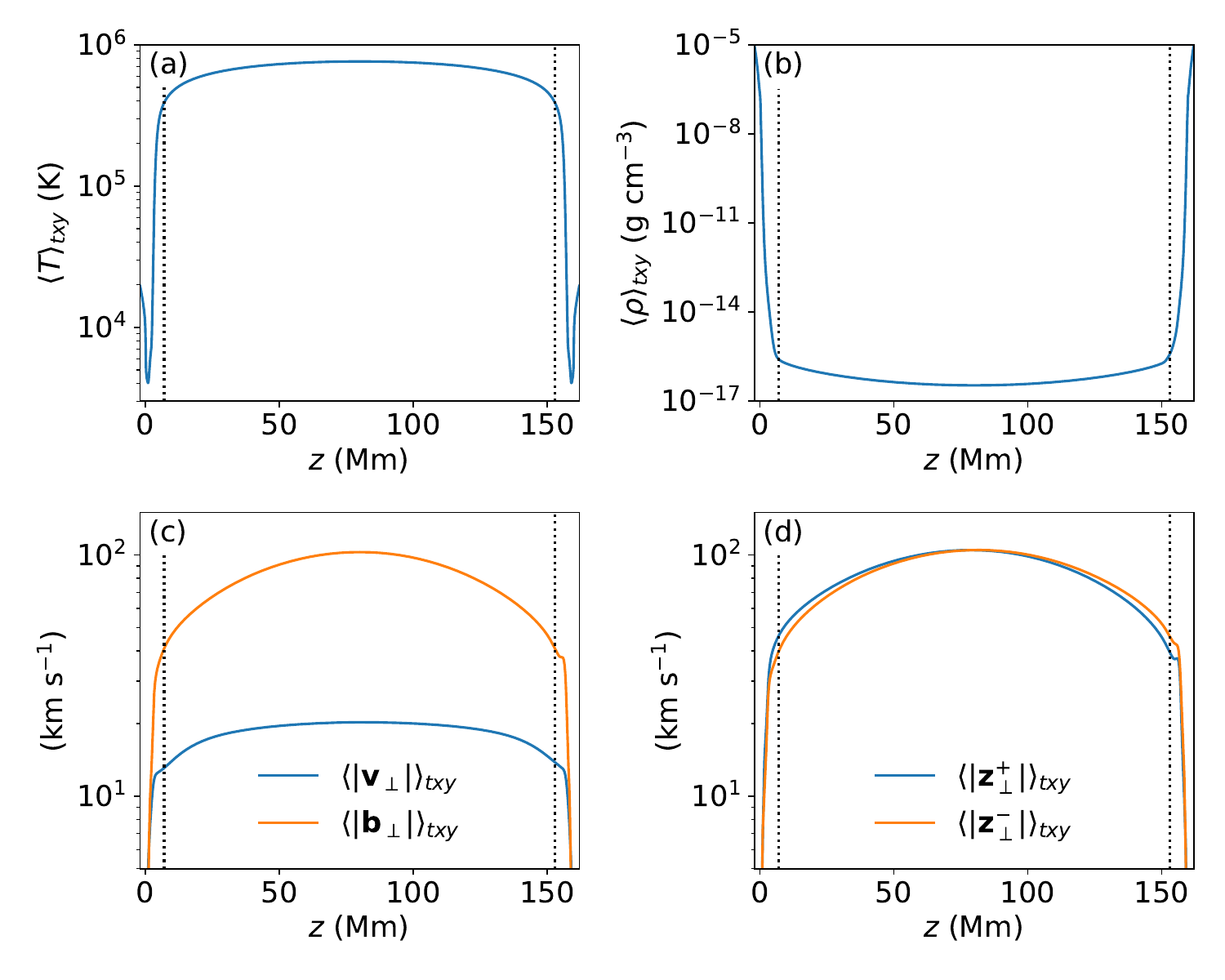} % 幅を10cmに指定
  \caption{
  Loop-aligned ($z$-directional) profiles of (a) temperature $\langle T \rangle_{txy}$, (b) mass density $\langle \rho \rangle_{txy}$, (c) horizontal velocity $\langle |\mathbf{v}_{\perp}| \rangle_{txy}$ and normalized magnetic field $\langle |\mathbf{b}_{\perp}| \rangle_{txy}$, and (d) Els{\"a}sser variables $\langle |\mathbf{z}^{\pm}_{\perp}| \rangle_{txy}$.
  \added{Dotted vertical lines indicate the coronal bases (z = 7 Mm and z = 153 Mm; Section 3.1). 
  Subscripts on angle brackets denote the coordinates over which each quantity is averaged.}
  }
  \label{fig:height_distribution}
\end{figure*}

\section{Numerical setup}\label{sec:numerical_setup}

We perform a three-dimensional radiative MHD simulation with the RAMENS\footnote{RAdiation Magnetohydrodynamics Extensive Numerical Solver} code \citep{Iijima_2016_PhDT, Iijima_2017_ApJ}, which solves compressible MHD equations including gravity, radiation, and thermal conduction in a Cartesian coordinate system. 
The governing equations are expressed in conservation form as:

\begin{align}
    & \frac{\partial \rho} {\partial t} + \nabla \cdot (\rho \boldsymbol{v} )  = 0, \\
    & \frac{\partial (\rho \boldsymbol{v})}{\partial t} 
      +\nabla \cdot \left[ \rho \boldsymbol{v} \boldsymbol{v}
      + \left( p+\frac{\boldsymbol{B}^2}{8\pi} \right) \boldsymbol{I}
      - \frac{\boldsymbol{B} \boldsymbol{B}}{4 \pi}  \right]
      = \rho \boldsymbol{g}, \\
    & \frac{\partial \boldsymbol{B}}{\partial t}+\nabla \cdot (\boldsymbol{vB}-\boldsymbol{Bv})=0, \\
    \label{eq:mhd_eqs}
    & \frac{\partial e}{\partial t} 
      + \nabla \cdot \left[
     \left(e+p+\frac{\boldsymbol{B}^2}{8\pi}\right)\boldsymbol{v} -\frac{1}{4\pi} \boldsymbol{B}(\boldsymbol{v}\cdot\boldsymbol{B}) \right] \\
    & =\rho \boldsymbol{g}\cdot \boldsymbol{v}+Q_{\mathrm{cnd}}+Q_{\mathrm{rad}} \nonumber , 
\end{align}

\noindent where $\rho$ is the mass density,  $\boldsymbol{v}$ is the gas velocity, $\boldsymbol{B}$ is the magnetic field, $e=e_{\rm int} + \rho \boldsymbol{v}^2/2 + \boldsymbol{B}^2/8\pi$ is the total energy density, $e_\mathrm{int}$ is the internal energy density,
$p$ is the gas pressure, $\boldsymbol{g}$ is the gravitational acceleration, and $\boldsymbol{I}$ is unit tensor.
$Q_\mathrm{cnd}$ and $Q_\mathrm{rad}$ denote the \added{heating/cooling} by thermal conduction and radiation, respectively. 
$Q_{\mathrm{cnd}}$ is the Spitzer-H{\"a}rm thermal conduction \citep{Spitzer_1953_PhRv}. 
The radiative term $Q_{\rm rad}$ is evaluated with a bridging law that combines optically thick and thin contributions \citep{Iijima_2016_PhDT}. 
The optically thick component is obtained by solving the gray radiative transfer equation under local thermodynamic equilibrium (LTE), while the optically thin component is computed from the CHIANTI atomic database \citep{Dere_1997_AAS, Landi_2012_ApJ} assuming coronal abundances. 
Because the loss function for the optically thin radiation is defined only for $T \geq 10^4 \ \mathrm{K}$, we extend it to lower temperatures following \citet{Goodman_2012_ApJ}. 
The equation of state is evaluated under LTE with the OPAL opacities \citep{Rogers_1996_ApJ}.

\medskip

Following the methodology outlined in \citet{Breu_2022_AA}, we model an entire coronal loop while neglecting its curvature.
The computational domain covers a horizontal extent of $9 \ \mathrm{Mm} \times 9 \ \mathrm{Mm}$ in the $xy$-plane and a vertical extent of $164 \ \mathrm{Mm}$ in the $z$–direction, ranging from $z=-2 \ \mathrm{Mm}$ to $z=162 \ \mathrm{Mm}$. 
The bottom ($z=-2 \ \mathrm{Mm}$) and top ($z=162 \ \mathrm{Mm}$) boundaries represent the upper convection zone. In this setup, the upper convection zone extends $2 \ \mathrm{Mm}$ beneath the optical depth unity surfaces, located at $z=0 \ \mathrm{Mm}$ and $z=160 \ \mathrm{Mm}$, respectively.
The gravitational acceleration is modeled according to a semi-circular loop approximation as

\begin{align}
    \boldsymbol{g} = -\frac{g \cos{\theta}}{(1 + h/R_{\mathrm{sun}})^2} \hat{\boldsymbol{z}},
\end{align}

\noindent 
where $g=2.74 \times 10^4 \ \mathrm{cm \ s^{-2}}$, $R_{\mathrm{sun}} = 6.96 \times 10^{10} \ \mathrm{cm}$, $\theta = z/r$, $h=r \sin{\theta}$, $r=L_z/\pi$, $L_z = 160 \ \mathrm{Mm}$, and $\hat{\boldsymbol{z}}$ is the unit vector in the $z$-direction.

\added{At the top and bottom boundaries, both of which represent the upper convection zone in this straightened-loop setup}, 
we apply open boundary conditions for outflows, while fixing the entropy of inflows to emulate convective energy input from the deeper convection zone \citep[for details, see][]{Vogler_2005_AA}.
Periodic boundary conditions are imposed in the $x$- and $y$-directions. 
The grid spacing is uniform at $60 \ \mathrm{km}$ in all three directions.

\medskip

While explicit resistivity and viscosity are not included in our calculations (see Equation~\eqref{eq:mhd_eqs}), heating still arises through numerical dissipation associated with discretization errors in the momentum and induction equations \citep{Matsumoto_2014_MNRAS}.
\added{We estimate the resulting local numerical heating rate per unit volume, 
denoted by $Q_{\rm num}$, following the method described in a previous study 
\citep{Kuniyoshi_2025_ApJ}.}
It is worth noting that this heating process remains physically meaningful because the small-scale dissipative structures originate from physical processes such as turbulent energy cascades. 
% Following the method described in previous studies \citep{Kuniyoshi_2025_ApJ}, we estimate the heating rate using the numerical dissipation rate, $Q_{\rm num}$.

\medskip

The initial state of the convection zones is specified by Model S \citep{ChristensenDalsgaard_1996_Sci}.
Above the surfaces, the atmosphere is extrapolated under hydrostatic equilibrium to establish a coronal temperature of $1 \ \rm MK$, and is permeated by a uniform vertical magnetic field of $B_z=6 \ \mathrm{G}$, consistent with typical quiet Sun coronal values \citep{Klimchuk_2006_SoPh}.
The system is first evolved for $3 \ \mathrm{hr}$ to allow the convection to relax, during which a conductive flux is applied at the coronal apex ($z=80 \ \rm Mm$) to maintain a coronal temperature above $1 \ \mathrm{MK}$.
This artificial flux is then switched off, and the simulation is continued for an additional $2.5 \ \mathrm{hr}$ so that the corona is heated self-consistently.
Our analysis focuses on the final $1.5 \ \mathrm{hr}$ of this phase, with snapshots saved every $2 \ \mathrm{s}$. In the following, we define $t=0 \ \mathrm{s}$ as the beginning of the analysis interval.

\section{Results}\label{sec:results}

\subsection{Simulation Overview}\label{subsec:simulation_overview}

\begin{figure}[t!]
  \centering
  \includegraphics[width=8.5cm]{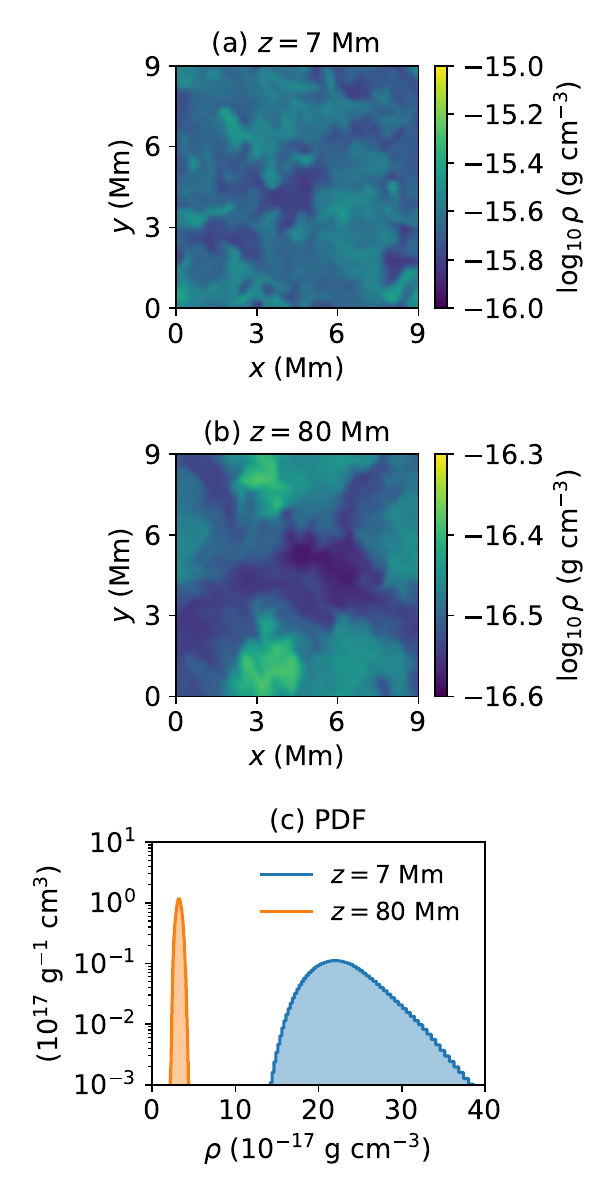} % 幅を10cmに指定
  \caption{
  (a,b): Snapshots of the mass density $\rho$ at $z = 7\ \rm Mm$ and $z = 80\ \rm Mm$ at $t = 2160\ \rm s$, respectively.
  (c): Probability density functions (PDFs) of $\rho$ at $z = 7\ \rm Mm$ and $z = 80\ \rm Mm$.
  The associated animation shows the temporal evolution of panels (a) and (b) over a period from $t=0 \ \rm{s}$ to $t=5400 \ \rm{s}$.
  \added{Panels (a) and (b) show $\log_{10}{\rho}$ as indicated by the colorbars; panel (c) shows the probability density function of $\rho$ itself.}
  }
  \label{fig:ro_pdf}
\end{figure}

The corona is self-consistently heated to $\sim 10^6~\mathrm{K}$ (Figure~\ref{fig:snapshots}a), consistent with quiet-Sun observations \citep{Warren_2009_ApJ, Brooks_2009_ApJ}.
Coronal mass is also self-consistently supplied through spontaneous chromospheric evaporation and jets, sustaining inhomogeneous density structures both along and across the background magnetic field (Figure~\ref{fig:snapshots}b and its associated animation).

The horizontal velocity ($\mathbf{v}_{\perp}=(v_x,v_y)$) and magnetic field ($\mathbf{B}_{\perp}=(B_x,B_y)$), which are perpendicular to the background magnetic field, should carry Alfv\'{e}nic perturbations.
However, their spatial structures lack the clear correlation expected from linear Alfv\'{e}n wave theory ($\mathbf{v}_{\perp} = \pm \mathbf{B}_{\perp}/\sqrt{4 \pi \rho}$; \citealt{Walen_1944_ArMAF, Priest_2014_textbook}), as shown in Figures~\ref{fig:snapshots}(c)--(d).
Nevertheless, propagating perturbations are visible, suggesting the presence of Alfv\'{e}nic waves (see the associated animation).

To better characterize Alfv\'{e}nic perturbations, we employ the Els\"{a}sser variables \citep{Elsasser_1950_PhRv},

\begin{equation}
    \mathbf{z}_{\perp}^{\pm} = \mathbf{v}_{\perp} \mp \mathbf{b}_{\perp},
\end{equation}

\noindent
where $\boldsymbol{b}_{\perp}=\mathbf{B}_{\perp}/\sqrt{4\pi\rho}$ is normalized magnetic field in velocity units.
$\boldsymbol{z}_{\perp}^{+}$ ($\boldsymbol{z}_{\perp}^{-}$) represents Alfv\'{e}nic waves propagating in the upward (downward) direction along the $z$-axis.
The Els\"{a}sser variables are widely used to describe Alfv\'{e}nic waves in both homogeneous and inhomogeneous plasmas \citep{Magyar_2019a_ApJ}, and indeed the propagation of Alfv\'{e}nic perturbations in the positive and negative $z$-directions is clearly visible (see the animation associated with Figures~\ref{fig:snapshots}(e)--(f)).

\medskip

We define the coronal segment as the region where the mean temperature $\langle T \rangle_{txy} \ge 0.4 \times 10^6~\mathrm{K}$ 
(Figure~\ref{fig:height_distribution}(a)), corresponding to $7~\mathrm{Mm} \le z \le 153~\mathrm{Mm}$, where angle brackets 
$\langle \rangle$ denote averaging over the directions indicated by the subscripts. 
Within this segment, the mean mass density 
$\langle \rho \rangle_{txy}$ lies in the range $10^{-17}$--$10^{-16}~\mathrm{g~cm^{-3}}$ (Figure~\ref{fig:height_distribution}(b)), consistent with previous observations \citep[e.g.,][]{Yang_2020_Sci}.

The mean horizontal velocity amplitude $\langle |\mathbf{v}_{\perp}| \rangle_{txy}$ in the corona lies in the range $10$--$20~\mathrm{km~s^{-1}}$ (Figure~\ref{fig:height_distribution}(c)), consistent with coronal observations of kink wave amplitudes \citep{McIntosh_2011_Natur, Thurgood_2014_ApJ, Morton_2025_nata, Morton_2019_NatAs, Morton_2025b_ApJ} and nonthermal velocities \citep{Chae_1998_ApJ, Hara_1999_ApJ, Asgari_2014_ApJ, Brooks_2016_ApJ}.
In contrast, the mean horizontal magnetic field $\langle |\mathbf{b}_{\perp}| \rangle_{txy}$ exceeds $\langle |\mathbf{v}_{\perp}| \rangle_{txy}$ by more than a factor of four, the origin of which is addressed in Section~\ref{subsec:temporal_evolution_of_coronal_aflvenic_waves}.

The mean amptlitudes of Els\"{a}sser variables $\langle |\mathbf{z}_{\perp}^{\pm}| \rangle_{txy}$ have comparable magnitudes near the coronal apex, while showing slight asymmetry near the coronal bases (Figure~\ref{fig:height_distribution}(d)). This indicates that Alfv\'{e}nic energy is injected comparably from both coronal bases.

\medskip

Alfv{\'e}nic waves in our simulation are primarily generated through the interaction between convective motions and the magnetic field at the photosphere, without any artificial wave driver. 
\added{The velocity power spectrum at the surface therefore emerges self-consistently from the photospheric convective motions, rather than being prescribed. 
Figure~\ref{fig:ph_psd} in Appendix~\ref{sec:appendix} shows a comparison between the photospheric and coronal horizontal velocity power spectra in frequency space.}
This is consistent with previous radiative MHD simulations \citep{Battaglia_2021_AA}.
Minor additional contributions may arise from interchange reconnection between unipolar and bipolar magnetic fields \citep[e.g.,][]{Isobe_2008_ApJ} and from mode conversion of acoustic waves \citep{Khomenko_2012_ApJ}, although quantifying their relative contributions is beyond the scope of the present study.

\subsection{Perpendicular density inhomogeneity}\label{subsec:density_inhomogeneity}

The coronal mass density is inhomogeneous perpendicular to the background magnetic field, exhibiting complex structures rather than coherent high-density bundles (Figures~\ref{fig:ro_pdf}(a)--(b)), consistent with previous numerical results \citep{Malanushenko_2022_ApJ, Kohutova_2024_AA}.
These density structures undergo periodic oscillations driven by Alfv\'{e}nic waves, as visible in the associated animation.
The mass density follows a positively skewed lognormal distribution (Figure~\ref{fig:ro_pdf}(c)), particularly near the coronal \added{bases} where dense material is continuously supplied from the chromosphere.
\added{Figure~\ref{fig:ro_pdf}(c) shows one coronal base ($z = 7~\mathrm{Mm}$) as a representative example; the same trend holds at the opposite base ($z = 153~\mathrm{Mm}$).}

\begin{figure}[t!]
  \centering
  \includegraphics[width=8cm]{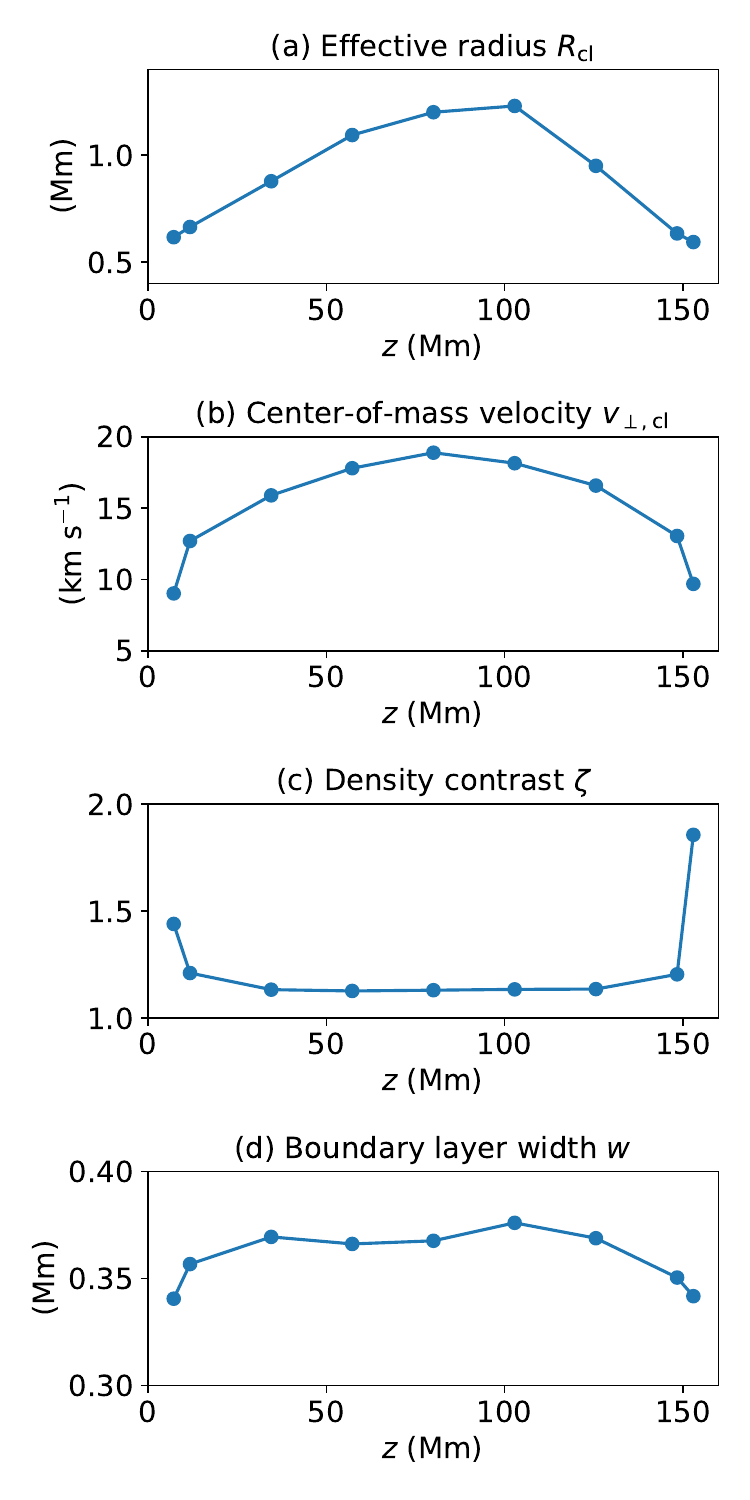} % 幅を10cmに指定
  \caption{
  Several mean properties of the density clumps as functions of $z$:
  (a) mean effective radius $R_{\rm cl}$,
  (b) center-of-mass velocity amplitude in the horizontal direction $v_{\perp,\rm cl}$, 
  (c) density contrast $\zeta$ and
  (d) width of boundary layer $w$.
  The standard error of the mean properties is negligible (at least $50$ times smaller than the mean values).
  }
  \label{fig:density_clump_properties}
\end{figure}

\medskip

In order to determine the typical scale and density of the inhomogeneity, we define density clumps as regions where $\rho(t,x,y,z)$ exceeds a threshold $\rho_{\rm th}(t,z)$, given by

\begin{equation}
    \rho_{\rm th}(t,z) = \exp{\left( \mu_{\rho}(t,z) + \sigma_{\rho}(t,z) \right)},
\end{equation}

\noindent
where $\mu_{\rho}(t,z)$ and $\sigma_{\rho}(t,z)$ are the mean and standard deviation of $\log{\rho}$ over the $xy$-plane at each coronal height.
For each clump, we measure its effective radius ($R_{\rm cl}$), center-of-mass velocity amplitude in the horizontal direction ($v_{\perp, \rm cl}$), and density contrast relative to the surrounding plasma ($\zeta$).
$R_{\rm cl}$ is estimated by approximating each clump as circular and computing its radius.
The clumps with radii smaller than five grid cells ($= 300 \ \rm km$) are discarded, as the dynamics of such small structures are likely to be affected by numerical discretization errors.
$v_{\perp, \rm cl}$ is defined as the root-mean-squared value of the horizontal velocity inside each clump.
The density contrast $\zeta$ is computed as the ratio of the average density inside the clump to that outside.

The mean radius over detected clumps $\langle R_{\rm cl} \rangle_{*}$ lies in the range $0.6$--$1.25\ \rm Mm$ (Figure~\ref{fig:density_clump_properties}(a)), consistent with observed coronal loop radii \citep[$0.5$--$10\ \rm Mm$,][]{Williams_2020_ApJ}. The effective radius increases with height, a trend inconsistent with magnetic-flux conservation; since the coronal magnetic field strength is nearly constant with height, flux conservation would require $R_{\rm cl}$ to remain nearly constant. Instead, we attribute this increase to turbulent mixing: as discussed in Section~\ref{subsec:nonlinearity}, the corona exhibits turbulent characteristics, and in such an environment, clumps interact and merge more frequently, forming larger aggregated structures.

The mean center-of-mass velocity $\langle v_{\perp,\rm cl} \rangle_{*}$ is $10$--$20\ \rm km\ s^{-1}$ (Figure~\ref{fig:density_clump_properties}(b)), comparable to $\langle |\mathbf{v}_{\perp}| \rangle_{txy}$ and consistent with the coronal observations noted above (Section~\ref{subsec:simulation_overview}).
The density contrast $\langle \zeta \rangle_{*}$ remains in the range $1.1$--$1.2$ at most altitudes except near the coronal \added{bases} (Figure~\ref{fig:density_clump_properties}(c)), consistent with observational estimates based on kink wave damping in quiet-Sun loops \citep[$\zeta \le 1.3$,][]{Morton_2021_ApJ},
while the larger values reported in previous observations 
(up to $\zeta \sim 10$) are associated with active-region 
loops \citep[e.g.,][]{Aschwanden_2003_ApJ}.
Near the coronal \added{bases} ($z=7\ \rm Mm$ and $z=153\ \rm Mm$), $\zeta$ increases to $1.5$--$2$, owing to chromospheric mass supply.

The width of the boundary layer of the density clumps is also estimated from $|\nabla_{\perp} \rho|$, which follows a skewed lognormal distribution similar to $\rho$.
Regions where $|\nabla_{\perp} \rho|$ exceeds one standard deviation of $\log{|\nabla_{\perp} \rho|}$ are identified as boundary layers, and their minor-axis length defines $w$.
Since individual layers cannot be clearly assigned to specific clumps, we instead average $w$ over all detected regions at each coronal height, yielding values of $340$--$370\ \rm km$ (Figure~\ref{fig:density_clump_properties}(d)).

\subsection{Temporal evolution of coronal Alfv{\'e}nic waves}\label{subsec:temporal_evolution_of_coronal_aflvenic_waves}

\begin{figure*}[t!]
  \centering
  \includegraphics[width=14cm]{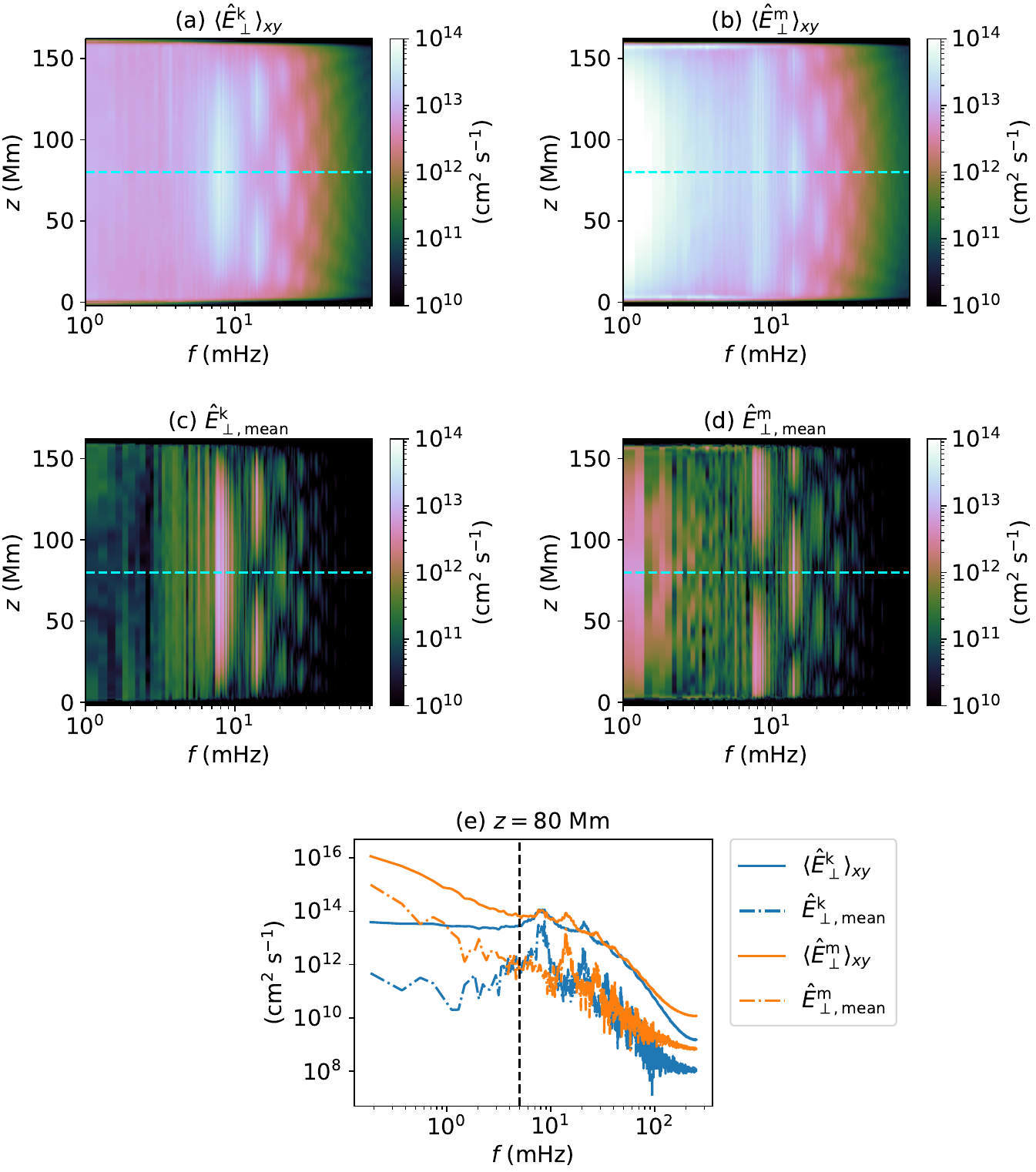} % 幅を10cmに指定
  \caption{
  (a,b): $f$--$z$ diagrams of the $xy$-averaged power spectral densities (PSDs) of the local horizontal velocity ($\langle \hat{E}^{\rm k}_{\perp} \rangle_{xy}$) and magnetic field ($\langle \hat{E}^{\rm m}_{\perp} \rangle_{xy}$).
  (c,d): $f$--$z$ diagrams of the mean-field PSDs of the horizontal velocity ($\hat{E}^{\rm k}_{\perp, \rm mean}$) and magnetic field ($\hat{E}^{\rm m}_{\perp, \rm mean}$). The cyan dashed lines show the position at $z=80\ \rm Mm$.
  (e): The local PSDs and mean-field PSDs, denoted as $\langle \hat{E}^{\rm k}_{\perp} \rangle_{xy}$, $\langle \hat{E}^{\rm m}_{\perp} \rangle_{xy}$, $\hat{E}^{\rm k}_{\perp, \rm mean}$, and $\hat{E}^{\rm m}_{\perp, \rm mean}$ at $z=80\ \rm Mm$.
  The black dashed line marks the position of $f=5\ \rm mHz$.
  For panels (a)--(d), the PSDs are computed on a downsampled grid (every 3 steps in $t$, 10 in $xy$, and 3 in $z$) due to memory constraints; consistency with full-resolution calculations has been verified at selected $z$ positions.
  }
  \label{fig:psd_omega_z_ekem_apex}
\end{figure*}

We compute the power spectral densities (PSDs) in frequency ($f$) space for the local and mean-field (over the $xy$-plane) horizontal velocity and magnetic field. 
The former (local PSDs) represent local motions within the simulated coronal loop such as torsional motions, whereas the latter (mean-field PSDs) represent bulk kink motions of the loop.

The local velocity and magnetic PSDs, $\hat{E}^{\rm k}_{\perp}(f,x,y)$ and $\hat{E}^{\rm m}_{\perp}(f,x,y)$, respectively, are defined as follows:

\begin{align}
    & \int \hat{E}^{\rm k}_{\perp}(f,x,y) \ df = \langle \boldsymbol{v}_{\perp}^2 (t,x,y) \rangle_t, \\
    & \int \hat{E}^{\rm m}_{\perp}(f,x,y) \ df = \langle \boldsymbol{b}_{\perp}^2(t,x,y) \rangle_t,
\end{align}

\noindent
Similarly, the mean-field PSDs, $\hat{E}^{\rm k}_{\perp, \rm mean}(f)$ and $\hat{E}^{\rm m}_{\perp, \rm mean}(f)$, respectively, are defined as follows:

\begin{align}
    & \int \hat{E}^{\rm k}_{\perp, \rm mean}(f) \ df = \langle \langle \boldsymbol{v}_{\perp} (t,x,y) \rangle_{xy}^2 \rangle_t \\
    & \int \hat{E}^{\rm m}_{\perp, \rm mean}(f) \ df = \langle \langle \boldsymbol{b}_{\perp} (t,x,y) \rangle_{xy}^2 \rangle_t.
\end{align}

The $f$--$z$ diagrams of the $xy$-averaged local PSDs (Figures~\ref{fig:psd_omega_z_ekem_apex}(a)--(b)), along with those of the corresponding mean-field PSDs (Figures~\ref{fig:psd_omega_z_ekem_apex}(c)--(d)), exhibit distinct structures characterized by multiple bands of enhanced power at regular frequency intervals.
These bands show nodes and antinodes elongated along the $z$-direction, indicative of resonant Alfv\'{e}nic modes in the corona.
For instance, the band around $f = 8.3~\mathrm{mHz}$ is consistent with the fundamental resonance frequency of a static coronal loop without gravitational stratification,

\begin{align}
    & f_1 \approx \frac{\langle c_{\rm A} \rangle_{txyz'}}{2 L_{\rm cor}},
\end{align}

\noindent 
where $c_{\rm A} = |\boldsymbol{B}|/\sqrt{4\pi\rho}$ is the Alfv\'{e}n speed, $z'$ denotes the vertical range of the coronal segment ($7\ \rm Mm \le z \le 153\ \rm Mm$), and $L_{\rm cor}$ is the mean coronal length.
With $\langle c_{\rm A} \rangle_{txyz'} = 2500~\mathrm{km~s^{-1}}$ and $L_{\rm cor} = 146~\mathrm{Mm}$, we obtain $f_1 = 8.6~\mathrm{mHz}$.
In practice, however, gravitational stratification causes the overtone ratios to deviate from pure harmonics \citep{Dymova_2006_AA, Morton_2011_AA}, leading to departures from this simple estimate.
This effect is addressed in Section~\ref{subsec:alfvenic_wave_resonance}.

The harmonic structures in the velocity PSDs (Figures~\ref{fig:psd_omega_z_ekem_apex}(a) and (c)) 
are consistent with those found in previous numerical models of coronal loops with Alfv\'{e}nic wave driving, 
for both kink \citep{Afanasyev_2020_AA} and torsional modes \citep{Tajfirouze_2025a_ApJ}.
The magnetic PSDs (Figures~\ref{fig:psd_omega_z_ekem_apex}(b) and (d)) show similar harmonic structures, 
but with nodes and antinodes shifted by $L_{\rm cor}/2$, 
corresponding to a $\pi/2$ phase shift most clearly visible in the bulk kink oscillation PSDs 
(Figures~\ref{fig:psd_omega_z_ekem_apex}(c) and (d)).

Notably, no clear node is present near the coronal apex in the fundamental eigenmode of the local magnetic PSD (Figure~\ref{fig:psd_omega_z_ekem_apex}(b)), suggesting that the expected $\pi/2$ phase shift is disrupted.
This may result from resonant absorption and KH instability \citep[e.g.,][]{Ionson_1978_ApJ, Terradas_2008_ApJ}, 
which converts bulk kink modes into incoherent local torsional modes, 
thereby disrupting the coherent structure of the magnetic fundamental eigenmode.
Since the bulk kink amplitude is smaller than that of local oscillations (Figure~\ref{fig:psd_omega_z_ekem_apex}(e)), 
this effect is likely significant only for the fundamental mode, which carries the highest power, and not for higher harmonics.

\medskip

\begin{figure}[t!]
  \centering
  \includegraphics[width=7.5cm]{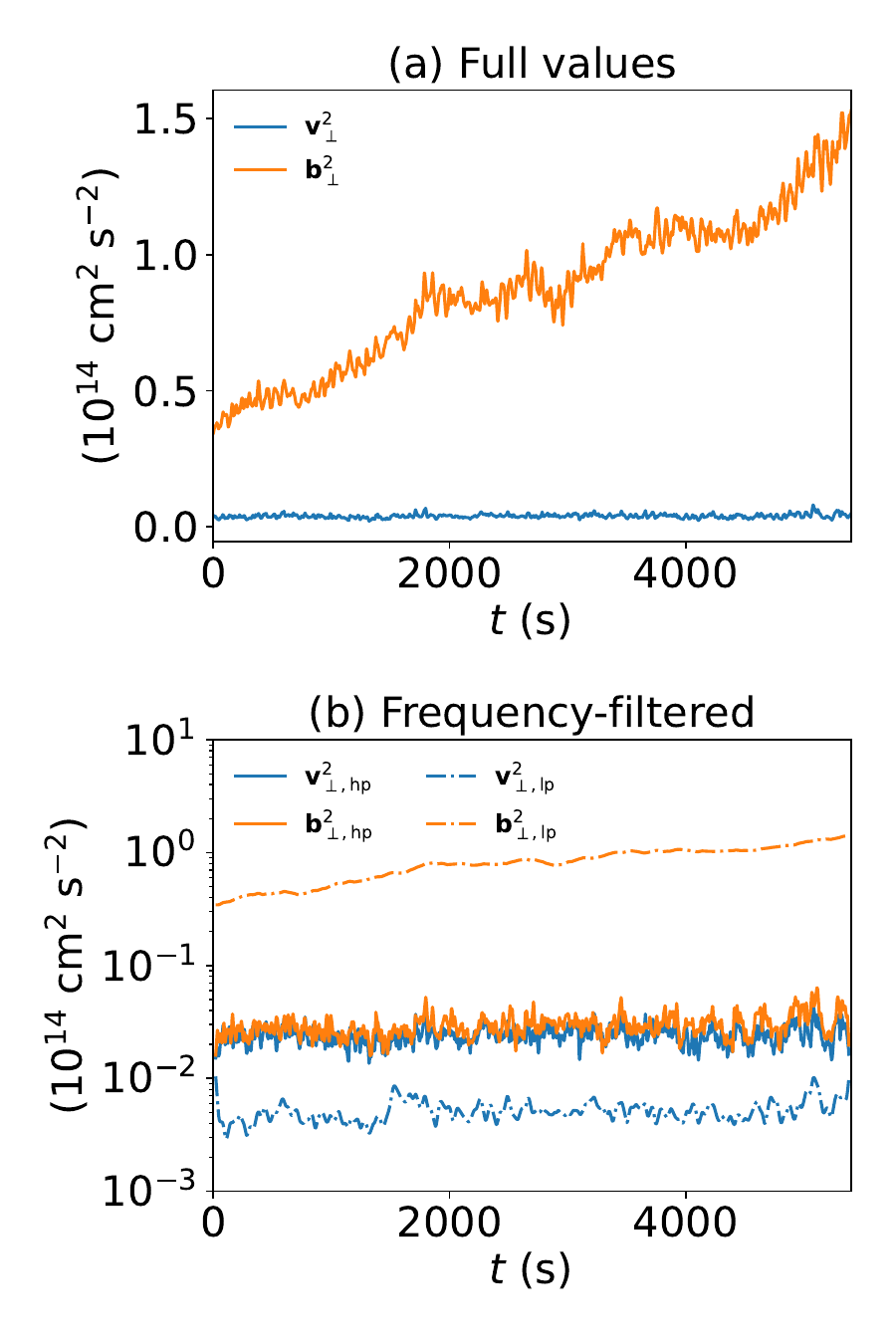} % 幅を10cmに指定
  \caption{
  The $xy$-averaged temporal evolution of the power of the horizontal velocity and the normalized magnetic field in velocity units at $z = 80~\rm Mm$.
  Panel (a) shows the unfiltered values ($\langle \boldsymbol{v}^2_{\perp} \rangle_{xy}$ and $\langle \boldsymbol{b}^2_{\perp} \rangle_{xy}$). 
  Panel (b) shows the low-pass–filtered components ($\langle \boldsymbol{v}^2_{\perp, \rm lp}\rangle_{xy}$ and $\langle \boldsymbol{b}^2_{\perp, \rm lp} \rangle_{xy}$) and high-pass–filtered components ($\langle \boldsymbol{v}^2_{\perp, \rm hp} \rangle_{xy}$ and $\langle \boldsymbol{b}^2_{\perp, \rm hp} \rangle_{xy}$).
  }
  \label{fig:ekem_xyav_apex}
\end{figure}

The PSDs show that magnetic power dominates over velocity power at lower frequencies ($f \lesssim 5~\mathrm{mHz}$), whereas the two are comparable at higher frequencies (Figure~\ref{fig:psd_omega_z_ekem_apex}(e)). We therefore define $f = 5~\mathrm{mHz}$ as the boundary between the low- and high-frequency regimes, with the Alfv\'{e}nic resonant modes located in the latter.
The time evolution of the energy components reveals the same frequency 
dependence.
The $xy$-averaged velocity power $\langle \boldsymbol{v}_{\perp}^2 
\rangle_{xy}$ remains nearly constant over time in the corona, while the 
magnetic power $\langle \boldsymbol{b}_{\perp}^2 \rangle_{xy}$ continues 
to increase (Figure~\ref{fig:ekem_xyav_apex}(a)).
Frequency filtering at $f = 5~\mathrm{mHz}$ reveals distinct behaviors between the low- and 
high-frequency regimes (Figure~\ref{fig:ekem_xyav_apex}(b)): magnetic 
power strongly exceeds velocity power and grows over time at low 
frequencies, whereas the two are comparable and nearly constant at high 
frequencies.
This frequency-dependent behavior indicates that magnetic energy is stored 
through continuous field-line tangling in a DC-like manner at low 
frequencies, whereas high frequencies are dominated by Alfv\'{e}nic waves 
in an AC-like process. 
Such behavior is consistent with previous numerical simulations 
based on the reduced MHD approximation \citep{Buchlin_2007_ApJ, Nigro_2008_ApJ, Verdini_2012_AA}, 
which assumes an incompressible plasma with a background magnetic field 
much stronger than the perturbations \citep{Strauss_1976_PhFl}.

\subsection{Correlation length of coronal Alfv{\'e}nic waves}
\label{subsec:correlation_length_of_the_aflvenic_waves}

\begin{figure}[t!]
  \centering
  \includegraphics[width=8.5cm]{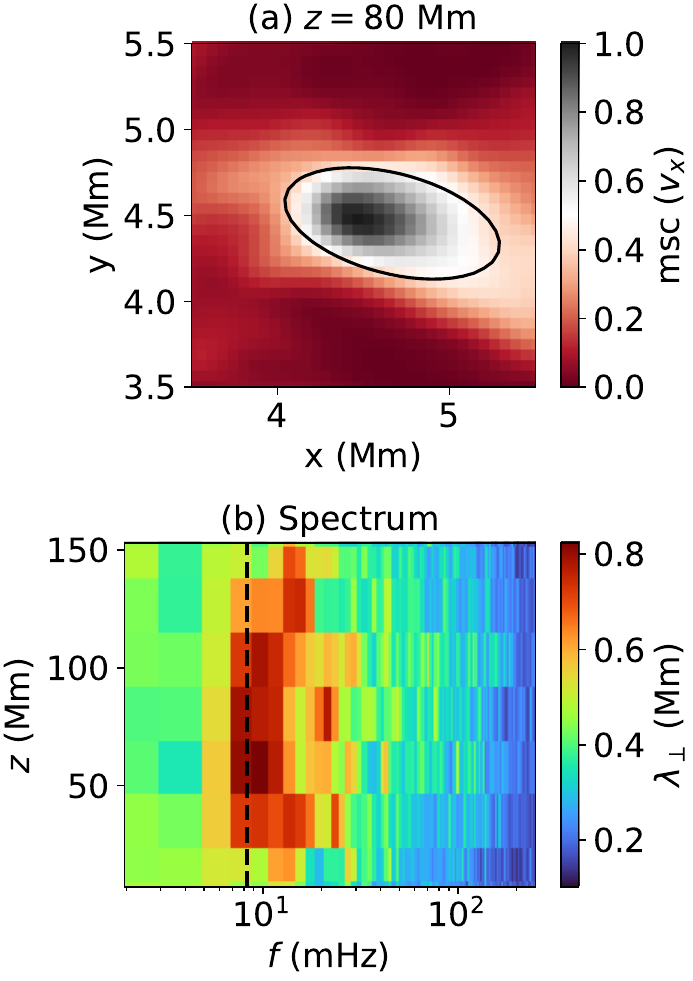} % 幅を10cmに指定
  \caption{
  (a): The mean squared coherence (MSC) map of $f = 8.3\ \rm mHz$ in the $xy$-plane at $z = 80\ \rm Mm$, derived from $v_x$. 
  The black contour is obtained from the fitted model and marks the region where the MSC exceeds $\exp(-0.5)$.
  (b): Spectrum of the correlation length $\lambda_{\perp}$ as a function of frequency $f$, stacked along the $z$-direction.
  }
  \label{fig:msc_corr_length}
\end{figure}

To quantify the horizontal length scale of Alfv\'{e}nic waves, we evaluate the mean squared coherence (MSC) of the horizontal velocity on the $xy$-plane at each coronal height, following \citet{Sharma_2023_NatAs}.
The MSC is calculated between the time series at pixel $\alpha$ and those at neighboring pixels $\beta$:

\begin{equation}
\label{eq:msc}
    \mathrm{MSC}(f) = \frac{|F_{\alpha \beta}(f)|^2}{F_{\alpha \alpha}(f) F_{\beta \beta}(f)},
\end{equation}

\noindent
where $F_{\alpha \beta}(f)$ is the cross-spectral density, and $F_{\alpha \alpha}(f)$ and $F_{\beta \beta}(f)$ are the corresponding auto-spectral densities. 
The Welch method \citep{Welch_1161901} is used to compute these spectra.

For $v_x$ and $v_y$, the MSC is computed with respect to a reference point at $x = y = 4.5\ \rm Mm$ within a $2 \times 2\ \rm Mm^2$ area (Figure~\ref{fig:msc_corr_length}(a)).
Fitting the MSC distribution with a two-dimensional Gaussian yields the correlation length $\lambda_{\perp} = \sqrt{2}\sigma_{\perp}$, where $\sigma_{\perp}$ is the average of the major-axis standard deviations of 
the Gaussian fits to $v_x$ and $v_y$ \citep{Sharma_2023_NatAs, Morton_2025a_ApJ, Tajfirouze_2025a_ApJ, Hahn_2025_ApJ}.
The resulting $f$--$z$ diagram of $\lambda_{\perp}$ shows a peak of $\lambda_{\perp} = \lambda_{\perp,1} \approx 0.8\ \rm Mm$ around the fundamental Alfv\'{e}nic resonance frequency $f = 8.3\ \rm mHz$ (Figure~\ref{fig:msc_corr_length}(b)).

\subsection{Nonlinearity}
\label{subsec:nonlinearity}

\begin{figure}[t!]
  \centering
  \includegraphics[width=8cm]{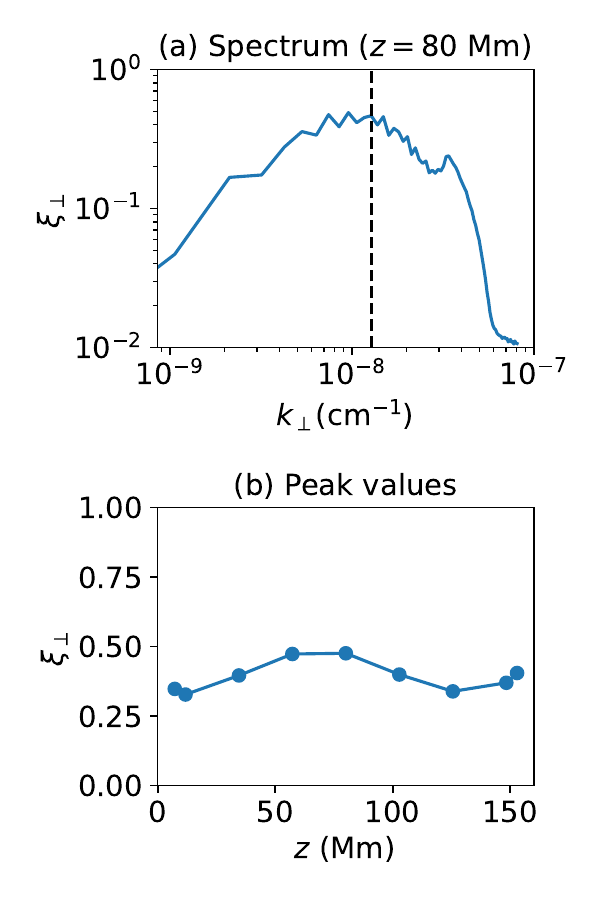} % 幅を10cmに指定
  \caption{
  (a) Spectrum of the nonlinear parameter $\xi_{\perp}$ as a function of horizontal wavenumber $k_{\perp}$ at $z = 80\ \rm Mm$. 
  The black dashed line indicates the wavenumber corresponding to $\lambda_{\perp, 1} = 0.8\ \rm Mm$ (see Figure~\ref{fig:msc_corr_length}(b)).
  (b) Peak values of $\xi_{\perp}$ over $k_{\perp}$ as a function of $z$.
  }
  \label{fig:nl_param}
\end{figure}

We quantify the degree of turbulence associated with Alfv{\'e}nic waves in the corona.
To achieve this, we employ the nonlinearity parameter, which is expressed as the ratio between the mean period of the Alfv{\'e}nic waves and the eddy turnover time, i.e., the characteristic timescale of the turbulent energy cascade perpendicular to the background magnetic field.

Following \citet{Goldreich_1995_ApJ}, we define the nonlinearity parameter $\xi_{\perp}(k_{\perp})$ at each horizontal wavenumber $k_{\perp} = \sqrt{k_x^2 + k_y^2}$, where $k_x$ and $k_y$ denote the wavenumbers in the $x$- and $y$-directions, respectively.
The parameter is expressed as

\begin{equation}
\xi_{\perp}(k_{\perp}) \equiv \frac{k_{\perp} \tilde{v}_{\perp}(k_{\perp})}{\tilde{f}(k_{\perp})},
\end{equation}

\noindent
where $\tilde{v}_{\perp}(k_{\perp})$ and $\tilde{f}(k_{\perp})$ are, at each wavenumber $k_{\perp}$, the frequency-integrated velocity 
amplitude and the power-weighted mean frequency, respectively. Both are derived from the frequency--wavenumber power spectrum 
$\hat{P}^{k}_{\perp}(k_{\perp}, f)$, which is defined such that

\begin{align}
    \int \hat{P}_{\perp}^{k} (k_{\perp}, f) \, dk_{\perp} \, df 
    = \langle \boldsymbol{v}_{\perp}^2 \rangle_{txy},
\end{align}

\noindent
and $\tilde{v}_{\perp}$ and $\tilde{f}$ are given as

\begin{align}
    & \tilde{v}^2_{\perp}(k_{\perp}) 
    = \frac{k_{\perp}}{\Delta k_{\perp}} 
      \int \hat{P}^{k}_{\perp}(k_{\perp}, f) \, df, \\
    & \tilde{f}(k_{\perp}) 
    = \frac{\int f \, \hat{P}^{k}_{\perp}(k_{\perp}, f) \, df}
           {\int \hat{P}^{k}_{\perp}(k_{\perp}, f) \, df},
\end{align}

\noindent
where $\Delta k_{\perp} = 2/\Delta x$ is set by the Nyquist limit.

The nonlinearity parameter $\xi_{\perp}$ peaks around $k_{\perp} = 10^{-8}\ \rm cm^{-1}$ (Figure~\ref{fig:nl_param}(a)), corresponding to the maximum correlation length $\lambda_{\perp}=\lambda_{\perp,1}$ (Figure~\ref{fig:msc_corr_length}(b)), and decreases at larger wavenumbers due to wave dissipation.
The peak values of $\xi_{\perp}$ along the loop-aligned direction lie in the range $0.25$--$0.5$ (Figure~\ref{fig:nl_param}(b)), showing that the spatial and temporal scales in the corona of this simulation can be associated with a state of weak turbulence. 
This result accounts for the coexistence of Alfv\'{e}nic resonance and turbulence \citep{Verdini_2012_AA, Tajfirouze_2025b_ApJ}.

\subsection{Alfv{\'e}nic wave resonance}\label{subsec:alfvenic_wave_resonance}

\begin{figure}[t!]
  \centering
  \includegraphics[width=8.5cm]{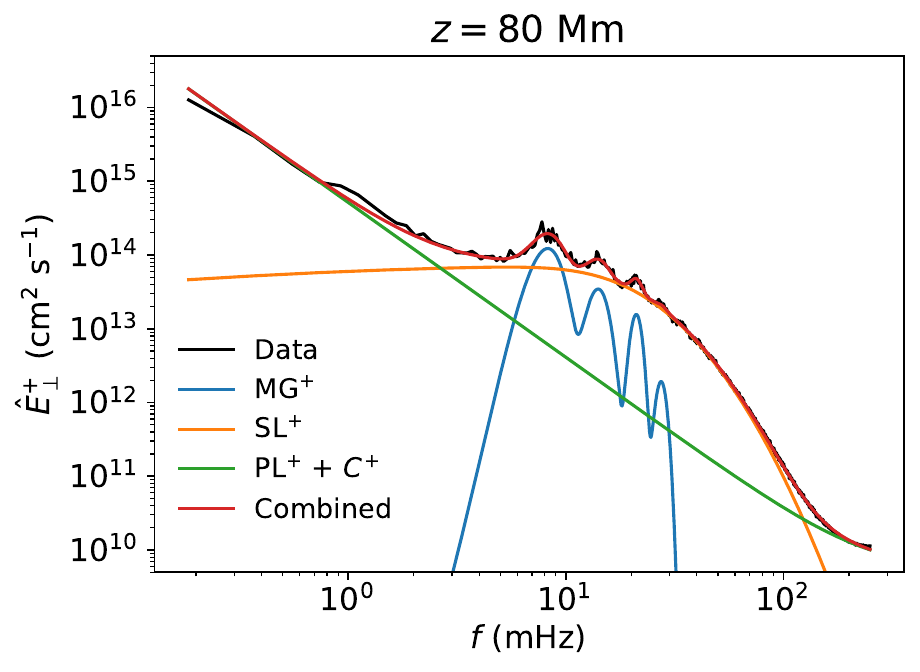} % 幅を10cmに指定
  \caption{
  The $xy$-averaged PSD of the upward Elsässer variable ($\langle \hat{E}_{\perp}^{+} \rangle_{xy}$) as a function of frequency $f$ is shown in black.
  Fitted results are overplotted: multiple Gaussian ($\mathrm{MG^{+}}$) in blue, skewed lognormal ($\mathrm{SL^{+}}$) in orange, power-law plus constant ($\mathrm{PL}^{+}+C^{+}$) in green, and the combined ($\mathrm{MG}^{+}+\mathrm{SL}^{+}+\mathrm{PL}^{+}+C^{+}$) in red.
  }
  \label{fig:zpt_peak_fitting}
\end{figure}

\begin{figure*}[t!]
  \centering
  \includegraphics[width=15cm]{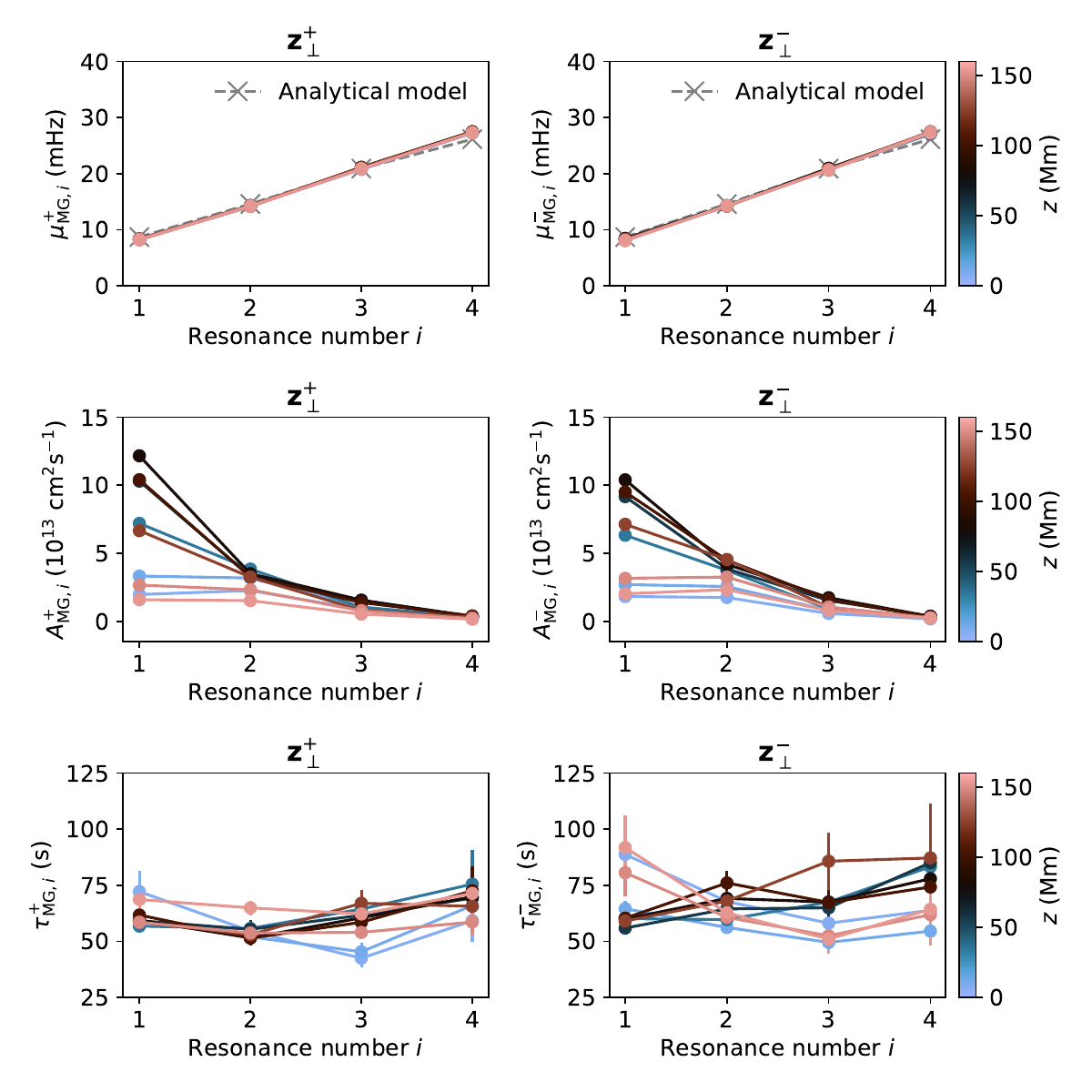} % 幅を10cmに指定
  \caption{
  Fitted resonant parameters for upward ($\mathbf{z}^{+}_{\perp}$) and downward ($\mathbf{z}^{-}_{\perp}$) Els{\"a}sser variables as a function of resonance number $i$ ($=1,2,3,4$). Line colors indicate the loop-aligned position $z$.
  (First row) Resonance frequencies $\mu^{\pm}_{\mathrm{MG},i}$, with the analytical model of Alfv{\'e}nic resonance in a 
  gravitationally stratified coronal loop \citep{Morton_2011_AA} shown as gray dashed lines.
  (Second row) Resonance amplitudes $A^{\pm}_{\mathrm{MG},i}$.
  (Third row) Resonance decay times $\tau^{\pm}_{\mathrm{MG},i}$.
  Error bars present fitting uncertainties, which are negligible except in the third column.
  }
  \label{fig:peak_properties}
\end{figure*}

To investigate the Alfv{\'e}nic wave resonance quantitatively, we calculate the PSDs of Els{\"a}sser variables in frequency space as 

\begin{align}
    \int \hat{E}^{\pm}_{\perp}(f,x,y) \ df = \langle (\boldsymbol{z}^{\pm}_{\perp})^2 (t,x,y) \rangle_t.
\end{align}

\noindent
The PSDs exhibit several peaks associated with Alfv\'{e}nic resonances (Figure~\ref{fig:zpt_peak_fitting}).

To characterize the Alfv{\'e}nic resonant modes, we model the PSDs of Els{\"a}sser variables as a combination of multiple Gaussian ($\rm MG^{\pm}$), skewed log-normal ($\rm SL^{\pm}$), power-law ($\rm PL^{\pm}$), and white-noise that is constant in frequency space ($C^{\pm}$), as follows:

\begin{equation}
    \langle \hat{E}_{\perp}^{\pm} \rangle_{xy} (f) = \mathrm{MG}^{\pm}(f) + \mathrm{SL}^{\pm}(f) + \mathrm{PL}^{\pm}(f) +  C^{\pm}.
\end{equation}

\noindent
$\rm MG^{\pm}$ represents the enhancements due to the resonance, whereas the other components constitute the baselines of the PSD.

The multiple Gaussian component $\rm MG^{\pm}$ is given by

\begin{equation}
    \mathrm{MG}^{\pm}(f) = \sum_{i=1}^{N} A^{\pm}_{\mathrm{MG},i} \exp{\left[ -\frac{(f - \mu^{\pm}_{\mathrm{MG}, i})^2}{2 (\sigma^{\pm}_{\mathrm{MG}, i})^2} \right]}, 
\end{equation}

\noindent
where, $A^{\pm}_{\mathrm{MG},i}$, $\mu^{\pm}_{\mathrm{MG},i}$ and $\sigma^{\pm}_{\mathrm{MG},i}$ are model parameters.
We model the first four resonances, i.e., $N=4$.
The skewed log-normal component $\rm SL^{\pm}$ is given by

\begin{equation}
    \begin{aligned}
        & \mathrm{SL}^{\pm}(f) = A^{\pm}_{\mathrm{SL}} \exp{\left[ -\frac{(\log_{10}f - \log_{10}\mu^{\pm}_{\mathrm{SL}})^2}{2 (\sigma^{\pm}_{\mathrm{SL}})^2} \right]} \\
        & \quad \times \left[ 1 + \mathrm{erf} \left( \frac{\alpha^{\pm}}{2} \frac{\log_{10}f - \log_{10}\mu^{\pm}_{\mathrm{SL}}}{\sigma^{\pm}_{\mathrm{SL}}} \right) \right],
    \end{aligned}
\end{equation}

\noindent
where $A^{\pm}_{\mathrm{SL}}$, $\mu^{\pm}_{\mathrm{SL}}$, $\sigma^{\pm}_{\mathrm{SL}}$ and $\alpha^{\pm}$ are model parameters. This is used to represent the transmission profile of upwardly propagating Alfv\'enic waves as first proposed by \citet{Soler_2019_ApJ} and observed in the corona by \citet{Morton_2025c_ApJ}.
The power-law component $\rm PL^{\pm}$ is expressed as 

\begin{equation}
    \mathrm{PL}^{\pm}(f) = A^{\pm}_{\mathrm{PL}} f^{-B^{\pm}_{\mathrm{PL}}},
\end{equation}

\noindent
where $A^{\pm}_{\mathrm{PL}}$ and $B^{\pm}_{\mathrm{PL}}$ are model parameters.
The white noise component $C^{\pm}$ is also a model parameter.
Following \citet{Morton_2025a_ApJ, Morton_2025c_ApJ}, all model parameters are fitted using a maximum-likelihood approach, with an example result shown in Figure~\ref{fig:zpt_peak_fitting}.

\begin{figure*}[t!]
  \centering
  \includegraphics[width=12cm]{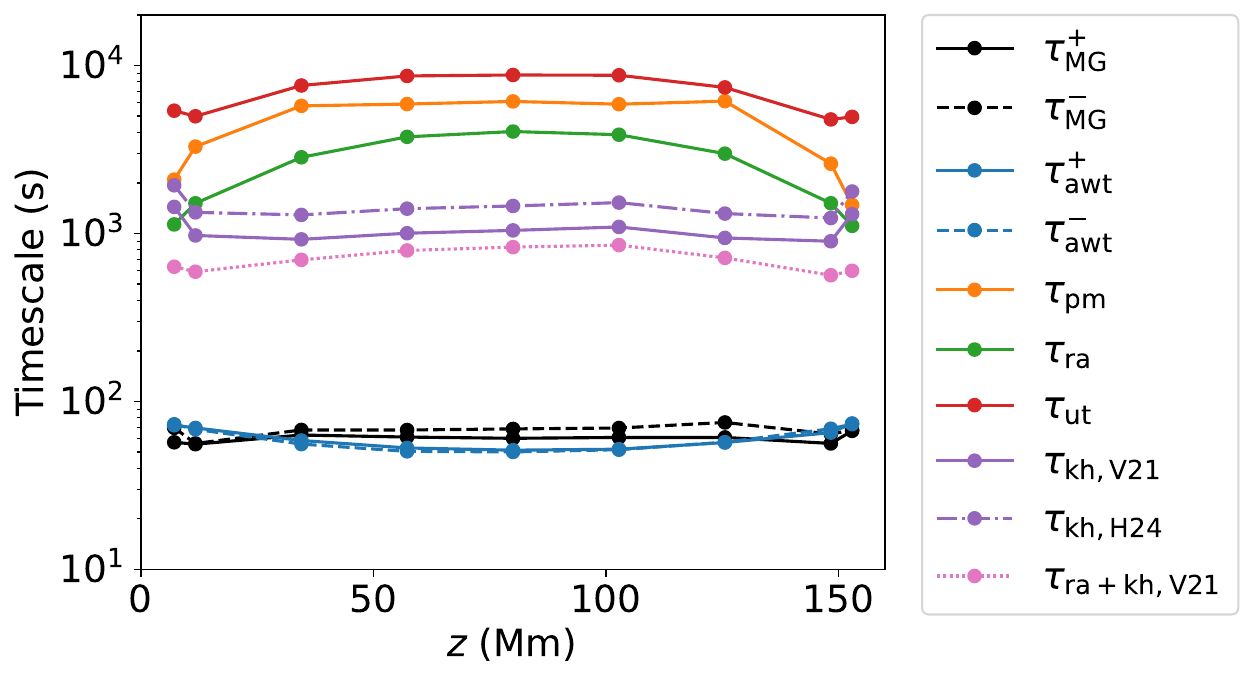} % 幅を10cmに指定
  \caption{
  Comparison of the decay time of Alfv{\'e}nic resonant modes and characteristic timescales of Alfv{\'e}nic wave damping mechanisms as a function of $z$.
  Black lines: mean decay time $\tau^{\pm}_{\mathrm{MG}}$ derived 
  from upward (solid) and downward (dashed) Els{\"a}sser variables.
  Blue lines: Alfv{\'e}n wave turbulence timescale $\tau_{\rm awt}^{\pm}$ for 
  upward (solid) and downward (dashed) Alfv{\'e}nic waves.
  Orange: phase mixing timescale $\tau_{\rm pm}$.
  Green: resonant absorption $\tau_{\rm ra}$.
  Red: uniturbulence timescale $\tau_{\rm ut}$.
  Purple solid: KH instability timescale $\tau_{\rm kh, V21}$ 
  \citep{VanDoorsselaere_2021a_ApJ, VanDoorsselaere_2021b_ApJ}.
  Purple dot-dashed: KH instability timescale $\tau_{\rm kh, H24}$ 
  \citep{Hillier_2024_ApJ}.
  Gray: combined timescale of resonant absorption and KH instability 
  $\tau_{\rm ra+kh, V21} = (1/\tau_{\rm ra} + 1/\tau_{\rm kh, V21})^{-1}$.
  }
  \label{fig:timescales}
\end{figure*}

\added{
Based on the fitted multiple-Gaussian components, we characterize each Alfv{\'e}nic resonant mode ($i=1,2,3,4$) by its resonance frequency ($\mu^{\pm}_{\mathrm{MG},i}$), amplitude ($A^{\pm}_{\mathrm{MG},i}$) and Gaussian width ($\sigma^{\pm}_{\mathrm{MG},i}$).
Following previous studies \citep{Hollweg_1984_ApJ, Soler_2021_ApJ}, 
we estimate the decay time of each resonant mode ($\tau^{\pm}_{\mathrm{MG}, i}$) 
from the full width at half maximum (FWHM) of the fitted Gaussian component,
$\Delta f^{\pm}_{\mathrm{MG},i}=2\sigma^{\pm}_{\mathrm{MG},i}\sqrt{2\ln 2}$.
The quality factor of each resonant mode is then defined as
$Q^{\pm}_{\mathrm{MG},i}=\mu^{\pm}_{\mathrm{MG},i}/\Delta f^{\pm}_{\mathrm{MG},i}$.
The quantity $Q^{\pm}_{\mathrm{MG},i}/2\pi$ corresponds to the ratio of the energy stored in the resonant mode to the energy lost during one oscillation period \citep{Feynman_1963_book}.
Since one oscillation period is $(\mu^{\pm}_{\mathrm{MG},i})^{-1}$, we estimate the decay time $\tau^{\pm}_{\mathrm{MG}, i}$ as
}

\begin{equation}
    \tau^{\pm}_{\mathrm{MG}, i}
    = \frac{Q^{\pm}_{\mathrm{MG},i}}{2 \pi}(\mu^{\pm}_{\mathrm{MG},i})^{-1}
    = \frac{1}{2 \pi \Delta f^{\pm}_{\mathrm{MG},i}}.
\end{equation}

The resonance properties ($\mu^{\pm}_{\mathrm{MG},i}$, $A^{\pm}_{\mathrm{MG},i}$, and $\tau^{\pm}_{\mathrm{MG},i}$) are summarized in Figure~\ref{fig:peak_properties}.
\added{Although $\mu^{\pm}_{\rm MG,i}$ is obtained from independent fits to the PSDs of the upward- and downward-propagating Els\"asser variables at each coronal height, the resulting values are nearly identical across coronal heights and between the two propagation directions. 
We therefore denote the common representative frequency of the $i$-th resonant mode as $\mu_{\mathrm{MG},i}$.}
In addition, they are consistent with the analytical model of Alfv\'{e}nic resonances in a gravitationally stratified coronal loop \citep{Morton_2011_AA}, with the best fit obtained by placing the coronal bases at $2\ \rm Mm$ above the surface rather than at $7\ \rm Mm$. This result indicates that the velocity node is located in the chromosphere.
The amplitude $A^{\pm}_{\mathrm{MG},i}$ decreases with resonance number $i$.
The decay time $\tau^{\pm}_{\mathrm{MG},i}$ shows no clear dependence on resonance number, in contrast to the frequency-dependent decay predicted by Alfv\'{e}nic resonance leakage \citep{Soler_2021_ApJ}, suggesting that the resonant modes are damped by dissipation rather than leakage.

\subsection{Time scale of Alfv{\'e}nic wave damping}
\label{subsec:time_scale_of__alfvenic_wave_dissipation}

We derive the characteristic damping timescales predicted by various analytical models (Alfv\'{e}n wave turbulence, phase mixing, resonant absorption, uniturbulence, and KH instability) and identify the most efficient mechanism by comparing them with the mean decay time of standing Alfv\'{e}nic waves $\tau^{\pm}_{\mathrm{MG}}=\langle \tau^{\pm}_{\mathrm{MG}, i} \rangle_{i}$.

\medskip
\noindent\textbf{Alfv\'{e}n wave turbulence:} Following \citet{Shoda_2018a_ApJ, Matsumoto_2021_MNRAS}, the Alfv\'{e}n wave turbulence timescale $\tau^{\pm}_{\rm awt}$ is estimated from the energy cascade timescale as

\begin{align}
    & \tau^{\pm}_{\rm awt} \approx \left \langle \left( \frac{|(\delta \boldsymbol{z}_{\perp, \rm{hp}}^{\mp} \cdot \nabla \boldsymbol{z}_{\perp, \rm{hp}}^{\pm})|}{|\boldsymbol{z}_{\perp, \rm{hp}}^{\pm}|} \right)^{-1} \right \rangle_{txy}, \\
    & \delta \boldsymbol{z}_{\perp, \rm{hp}}^{\pm} = \boldsymbol{z}_{\perp, \rm{hp}}^{\pm} - \langle \boldsymbol{z}_{\perp, \rm{hp}}^{\pm} \rangle_{xy}.
\end{align}

\noindent
It is worth noting that we use the high-pass-filtered ($f>5 \ \rm mHz$) Els{\"a}sser variables $\boldsymbol{z}^{\pm}_{\perp,{\rm hp}}$, in order to exclude the influence of the magnetic field line buildup occurring in a DC-like manner (see Section~\ref{subsec:temporal_evolution_of_coronal_aflvenic_waves}).

\medskip

\noindent\textbf{Phase mixing:} The characteristic length scale of phase mixing $l_{\rm pm}$ \citep{Mann_1995_JGR, Kaneko_2015_ApJ, Raes_2017_AA} is given by

\begin{equation}
    l_{\rm pm} = \frac{2 L_{\rm cor}}{t |\nabla_{\perp} c_{\rm A}|},
\end{equation}

\noindent
where $L_{\rm cor}$ is the mean coronal length and $c_{\rm A}$ is the Alfv{\'e}n speed, as defined in Section~\ref{subsec:temporal_evolution_of_coronal_aflvenic_waves}.
Since $l_{\rm pm}$ decreases with time, we define the phase-mixing timescale $\tau_{\rm pm}$ as the time required for $l_{\rm pm}$ to decrease from the correlation length of the fundamental Alfv{\'e}nic resonance $\lambda_{\perp, 1}$ (see Section~\ref{subsec:correlation_length_of_the_aflvenic_waves}) to the dissipation scale $l_{\rm dis} = 300\ \rm km$ (five grid spacings), below which numerical dissipation becomes non-negligible.
$\tau_{\rm pm}$ is expressed as follows:

\begin{equation}
  \tau_{\rm pm} = \left \langle \frac{2L_{\rm cor}}{|\nabla_{\perp} c_{\rm A}|} \left( \frac{1}{l_{\rm dis}} - \frac{1}{\lambda_{\perp, 1}} \right) \right \rangle_{txy}.
\end{equation}

\medskip

\noindent\textbf{Resonant absorption:} 
We define $\tau_{\rm ra}$ as the damping timescale of fundamental standing kink mode via resonant absorption, 
which can be expressed using the mean clump properties: the effective radius $R_{\rm cl}$, boundary layer width $w$, and density contrast $\zeta$ (see Section~\ref{subsec:density_inhomogeneity}).
Following \citet{Terradas_2010_AA}, $\tau_{\rm ra}$ is expressed as

\begin{equation}
  \tau_{\rm ra} = \mu_{\rm MG, 1}^{-1} \cdot \left \langle \frac{2}{\pi} \frac{R_{\rm cl}}{w} \frac{\zeta + 1}{\zeta - 1} \right \rangle_{*},
\end{equation}

\noindent \added{where $\mu_{\rm MG,1}(=8.3 \ \rm{mHz})$ is the common representative frequency of the fundamental resonant mode defined in Section~\ref{subsec:alfvenic_wave_resonance}, so that $\mu_{\rm MG,1}^{-1}$ gives its period.}

\medskip
\noindent\textbf{Uniturbulence:} The damping timescale due to uniturbulence $\tau_{\rm ut}$ is derived analytically by \citet{VanDoorsselaere_2020_ApJ} using the clump properties $R_{\rm cl}$, $\zeta$ and center-of-mass velocity $v_{\perp,\rm cl}$ (see Section~\ref{subsec:density_inhomogeneity}):

\begin{equation}
    \tau_{\rm ut} = \left\langle 2\sqrt{5\pi} \frac{R_{\rm cl}}{v_{\perp,\rm cl}} \frac{\zeta + 1}{\zeta - 1} \right\rangle_{*}.
\end{equation}

\medskip
\noindent\textbf{KH instability:} The damping timescale due to KH instability $\tau_{\rm kh}$ is evaluated using the clump properties $R_{\rm cl}$, $v_{\perp,\rm cl}$, and $\zeta$. 
Two independent analytical expressions are available: 
\citet{VanDoorsselaere_2021a_ApJ, VanDoorsselaere_2021b_ApJ} (taking the erratum into account) give

\begin{equation}
    \tau_{\rm kh, V21} = \left \langle 40 \sqrt{\pi} \frac{R_{\rm cl}}{v_{\perp, \rm cl}} \frac{\zeta + 1}{\sqrt{\zeta^2 -2 \zeta + 97}} \right \rangle_{*}.
\end{equation}

\noindent
and \citet{Hillier_2024_ApJ} independently derives

\begin{equation}
    \tau_{\rm kh, H24} = \left \langle \frac{\pi R_{\rm cl}/2}{0.3 v_{\perp, \rm{cl}}} \frac{(\sqrt{\zeta}+1)^2}{\zeta^{1/4}} \right \rangle_{*}.
\end{equation}

\noindent
$\tau_{\rm kh, V21}$ and $\tau_{\rm kh, H24}$ are in good agreement, with $\tau_{\rm kh, H24}$ being slightly longer by a factor of $1.3$--$1.4$ (Figure~\ref{fig:timescales}).

\medskip

The characteristic timescales of each Alfv\'{e}nic wave damping mechanism are summarized in Figure~\ref{fig:timescales}.
$\tau^{\pm}_{\rm awt}$ falls within $40$--$90\ \rm s$, while $\tau_{\rm pm}$, $\tau_{\rm ra}$, $\tau_{\rm ut}$, $\tau_{\rm kh, V21}$, and $\tau_{\rm kh, H24}$ are more than an order of magnitude larger, presenting that Alfv\'{e}n wave turbulence is the most efficient damping mechanism. The large timescales of the other mechanisms reflect the small density contrast $\zeta \approx 1.2$ in our simulation.
Even when resonant absorption enhances other mechanisms \citep{Antolin_2019_FrP, VanDoorsselaere_2021a_ApJ, VanDoorsselaere_2021b_ApJ}, the shortest combined timescale $\tau_{\rm ra+kh, V21} = (1/\tau_{\rm ra} + 1/\tau_{\rm kh, V21})^{-1}$ remains $8$--$16$ times longer than $\tau^{\pm}_{\rm awt}$.
The mean decay time $\tau^{\pm}_{\mathrm{MG}}$ is consistent with $\tau^{\pm}_{\rm awt}$ within a factor of $1.3$ (Figure~\ref{fig:timescales}), confirming that the resonant modes are primarily damped by Alfv\'{e}n wave turbulence.

\subsection{Coronal heating rate of Alfv{\'e}n wave turbulence}
\label{subsec:coronal_heating_rate}

\begin{figure}[t!]
  \centering
  \includegraphics[width=8.5cm]{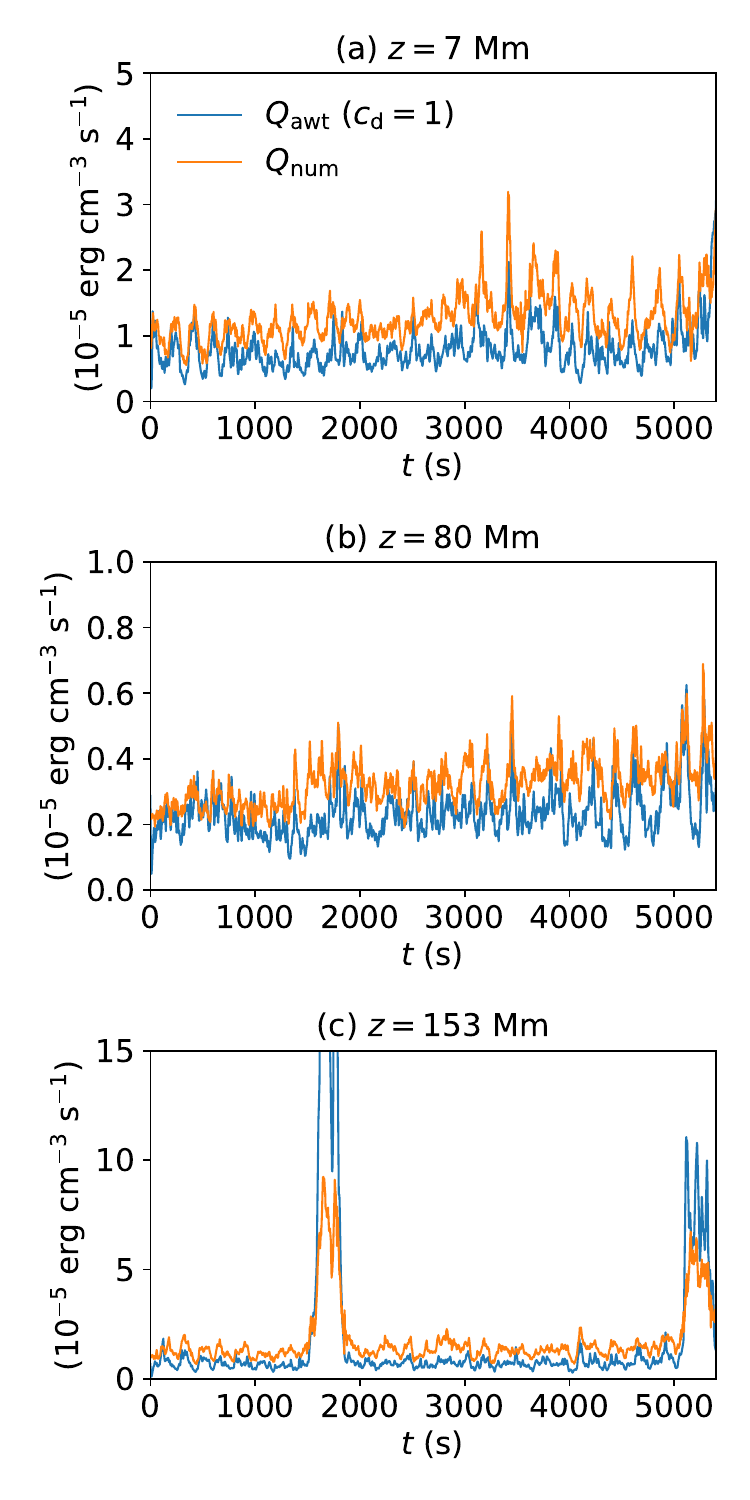} % 幅を10cmに指定
  \caption{
  Time evolution of the $xy$-averaged phenomenological heating rate from Alfv{\'e}n wave turbulence model $\langle Q_{\rm awt} \rangle_{xy}$ and numerical heating rate $\langle Q_{\rm num} \rangle_{xy}$ at (a) $z = 7\ \rm Mm$, (b) $z = 80\ \rm Mm$, and (c) $z = 153\ \rm Mm$.
  \added{Here, $Q_{\rm awt}$ is evaluated with the fixed value $c_{\rm d} = 1$ for reference.}
  }
  \label{fig:qa123}
\end{figure}

\begin{figure}[t!]
  \centering
  \includegraphics[width=8.5cm]{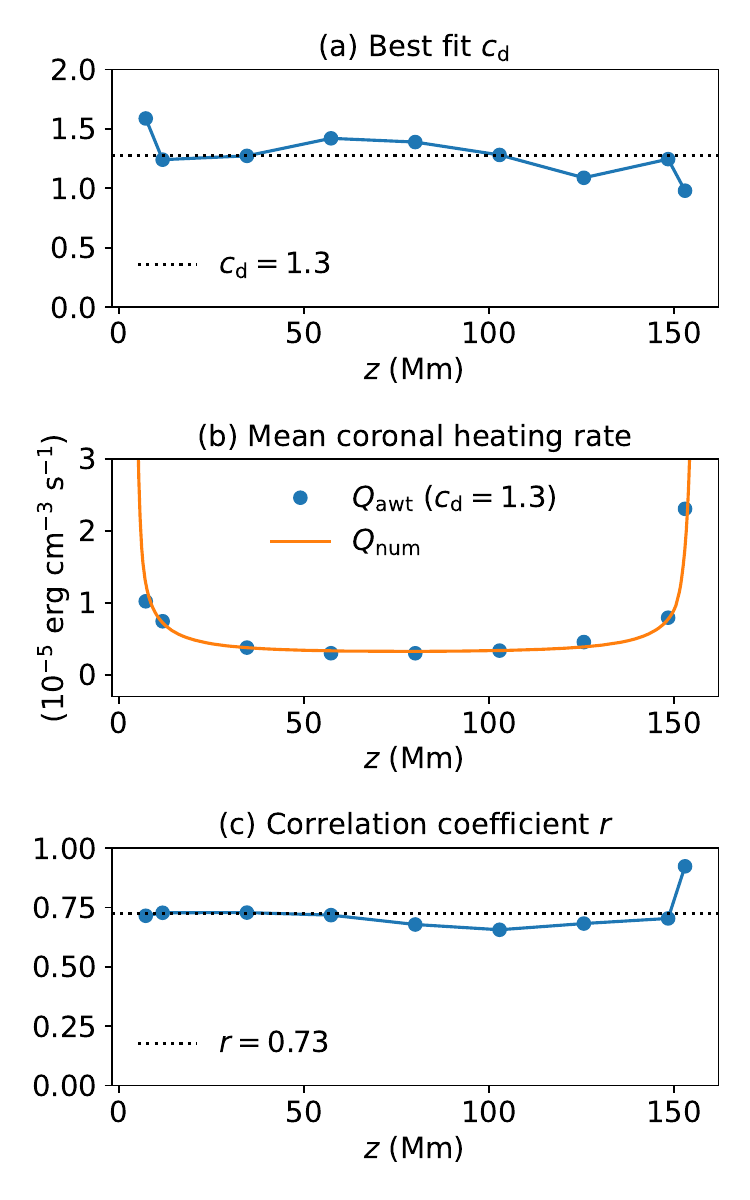} % 幅を10cmに指定
  \caption{
  (a) The value of $c_{\rm d}$ that provides the best fit of the phenomenological estimate to the numerical heating rate.
  (b) Horizontally and temporally averaged Alfv{\'e}n wave turbulence heating rate, $\langle Q_{\rm awt} \rangle_{txy}$, and numerical heating rate, $\langle Q_{\rm num} \rangle_{txy}$.
  (c) Correlation coefficient between the time series of $\langle Q_{\rm awt} \rangle_{xy}$ and $\langle Q_{\rm num} \rangle_{xy}$, denoted as $r$.
  \added{In panels (b) and (c), $Q_{\rm awt}$ is evaluated with the fixed value $c_{\rm d} = 1.3$, the best-fit value (panel (a)).}
  All quantities are shown as functions of $z$.
  }
  \label{fig:heating_rate_comparison}
\end{figure}

To further support the dominant role of Alfv\'en wave turbulence, 
we compare the coronal heating rate within the simulation to that 
expected from a phenomenological Alfv\'en wave turbulence model.
Following \citet{Hossain_1995_PhFl} and \citet{Matthaeus_1999_ApJ}, 
we estimate the phenomenological heating rate $Q_{\rm awt}$ as

\begin{equation}
\label{eq:awt_heating_rate}
    Q_{\rm awt} = c_{\rm d}\rho \frac{|\mathbf{z}_{\perp, \rm{hp}}^{+}| (\boldsymbol{z}_{\perp, \rm{hp}}^{-})^2 + |\mathbf{z}_{\perp, \rm{hp}}^{-}| (\boldsymbol{z}_{\perp, \rm{hp}}^{+})^2}{4 \lambda_{\perp}},
\end{equation}

\noindent
where $c_{\rm d}$ is a dimensionless parameter.
Here, consistent with the timescale estimation in Section~\ref{subsec:time_scale_of__alfvenic_wave_dissipation}, we use the high-pass-filtered Els\"{a}sser variables $z^{\pm}_{\perp,\rm hp}$ to suppress the DC-like contribution from magnetic field-line buildup (Section~\ref{subsec:temporal_evolution_of_coronal_aflvenic_waves}).
We take $\lambda_{\perp} = \lambda_{\perp,1}$ as the correlation length of the fundamental Alfv\'{e}nic resonance (Figure~\ref{fig:msc_corr_length}(b)) and $c_{\rm d} = 1$.
By construction, $Q_{\rm awt}$ does not include explicit information about the wavenumber spectrum in the cascading directions (primarily horizontal), so we consider the $xy$-averaged values of $Q_{\rm awt}$ in the following analysis.

$Q_{\rm awt}$ is evaluated at each coronal height and compared with the numerical dissipation rate $Q_{\rm num}$ in our simulation. 
\added{In the following comparisons, we use the averaged quantities indicated by the angle-bracket subscripts, rather than local grid-point values.}
$\langle Q_{\rm awt} \rangle_{xy}$ and $\langle Q_{\rm num} \rangle_{xy}$ exhibit qualitatively similar amplitudes and temporal variations (Figure~\ref{fig:qa123}).
At a coronal base, two pronounced enhancements in the coronal heating rate are visible (Figure~\ref{fig:qa123}(c)), corresponding to strong Alfv\'{e}nic wave pulses triggered by uni-directional magnetic field twisting associated with small-scale photospheric swirls in intergranular lanes \citep{Kuniyoshi_2023_ApJ, Kuniyoshi_2024_ApJ, Kuniyoshi_2025_ApJ}.
These impulsive events are not examined further here; we focus on the global Alfv\'{e}nic wave dynamics.

We further investigate the quantitative agreement between $Q_{\rm awt}$ and $Q_{\rm num}$.
The best-fit value of $c_{\rm d}$ is defined as the ratio $\langle Q_{\rm num} \rangle_{txy} / \langle Q_{\rm awt}(c_{\rm d}=1) \rangle_{txy}$, and ranges from $1$ to $1.6$ with a mean of $1.3$ (Figure~\ref{fig:heating_rate_comparison}(a)).
With $c_{\rm d} = 1.3$, $\langle Q_{\rm awt} \rangle_{txy}$ and $\langle Q_{\rm num} \rangle_{txy}$ show good agreement throughout the corona (Figure~\ref{fig:heating_rate_comparison}(b)).
The validity of $c_{\rm d} = 1.3$ is discussed in Section~\ref{subsec:phenomenological_heating_rate_from_alfven_wave_turbulence}.
In addition, the Pearson correlation coefficients between the temporal variations of $\langle Q_{\rm awt} \rangle_{xy}$ and $\langle Q_{\rm num} \rangle_{xy}$ range from $r = 0.66$ to $0.92$, with a height-averaged value of $r = 0.73$ (Figure~\ref{fig:heating_rate_comparison}(c)), further confirming the agreement between the phenomenological and numerical heating rates.

\subsection{Contribution of Alfv{\'e}nic wave induced shock heating}
\label{subsec:shock_heating}

\begin{figure}[t!]
  \centering
  \includegraphics[width=8cm]{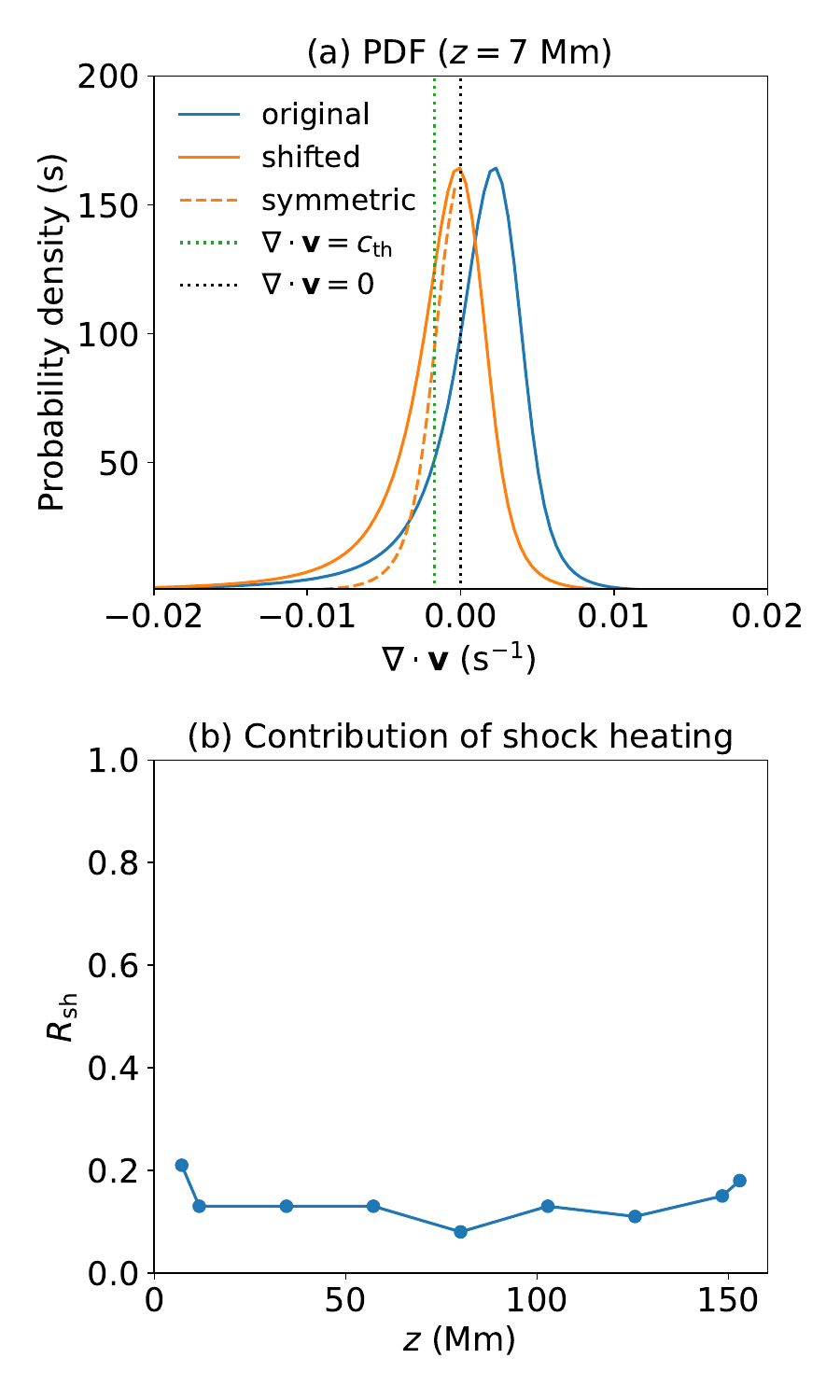} % 幅を10cmに指定
  \caption{
  (a) Blue solid line: the PDF of $\nabla \cdot \mathbf{v}$ at $z = 7\ \rm Mm$. 
  Orange solid line: the shifted PDF of $\nabla \cdot \mathbf{v}$, with its peak aligned at $\nabla \cdot \mathbf{v} = 0$. 
  Orange dashed line: expected symmetric component of the shifted PDF for $\nabla \cdot \mathbf{v} < 0$ with respect to $\nabla \cdot \mathbf{v} = 0$. 
  Dotted lines indicate $\nabla \cdot \mathbf{v} = 0$ (black) and $\nabla \cdot \mathbf{v} = c_{\rm th}$ (green).
  (b) The $xy$-averaged shock heating contribution to the total thermal energy input, $\langle R_{\rm sh} \rangle_{xy}$, as a function of $z$.
  }
  \label{fig:compressible_rate}
\end{figure}

We assess the contribution of Alfv\'{e}nic wave induced shock heating to the entire coronal heating.
At each coronal height, shocks are identified by locating regions of strong compression using the velocity divergence $\nabla \cdot \mathbf{v}$ \citep{Wang_2020_ApJ},

\begin{equation}
    \nabla \cdot \mathbf{v} < c_{\rm th},
\end{equation}

\noindent where $c_{\rm th}$ is a threshold determined as follows.
For linear waves, the PDF of $\nabla \cdot \mathbf{v}$ is symmetric about zero, whereas shocks produce an asymmetric tail toward negative values (Figure~\ref{fig:compressible_rate}(a)).
We therefore estimate the linear wave contribution from the positive side and set $c_{\rm th}$ where the negative tail exceeds twice the linear contribution, ensuring that only genuine shock fronts are retained.
Near the coronal bases, the PDFs are slightly shifted toward positive values due to upflows of the chromospheric mass supply. 
This bias is removed prior to the analysis by shifting each distribution so that its peak lies at $\nabla \cdot \mathbf{v} = 0$.

Following \citet{Matsumoto_2016_MNRAS}, we estimate the shock heating contribution by dividing the analysis interval ($1.5~\mathrm{hr}$) at each grid point into a shock duration,

\begin{equation}
    \Omega_{\rm sh}(x,y,z) = \{\, t \mid \nabla \cdot \mathbf{v} < c_{\rm th} \,\},
\end{equation}

\noindent and a no-shock duration,

\begin{equation}
    \Omega_{\rm ns}(x,y,z) = \{\, t \mid \nabla \cdot \mathbf{v} \ge c_{\rm th} \,\}.
\end{equation}

\noindent Unlike \citet{Matsumoto_2016_MNRAS}, who adopted $c_{\rm th} = 0$, we set $c_{\rm th} < 0$ to better isolate shocks from linear wave compression (Figure~\ref{fig:compressible_rate}(a)).
The thermal energy supplied by shock heating and incompressible heating (primarily Alfv\'{e}n wave turbulence) over the analysis interval is then

\begin{align}
    & e_{\rm int, sh}(x,y,z) = \int_{\Omega_{\rm sh}} Q_{\rm num}\ dt, \\
    & e_{\rm int, ns}(x,y,z) = \int_{\Omega_{\rm ns}} Q_{\rm num}\ dt,
\end{align}

\noindent and the fractional contribution of shock heating to the total thermal energy input is

\begin{equation}
    R_{\rm sh}(x,y,z) = \frac{e_{\rm int, sh}}{e_{\rm int, sh} + e_{\rm int, ns}}.
\end{equation}

The shock heating contribution $\langle R_{\rm sh} \rangle_{xy}$ is 
approximately $10\%$ throughout most of the corona, except near the 
coronal \added{bases}, where it reaches $\sim 20\%$ (Figure~\ref{fig:compressible_rate}(b)).
This indicates that Alfv\'{e}nic wave induced shock heating plays only a minor role in the overall coronal heating.
This is broadly consistent with \citet{Matsumoto_2016_MNRAS}, who report a shock heating contribution below $10\%$ above the magnetic canopy, though the elevated value near the coronal \added{bases} in our simulation may reflect strong Alfv\'{e}nic wave pulses driven by uni-directional photospheric swirls (Figure~\ref{fig:qa123}(c)).
Such pulses have larger amplitudes than those driven by random granular buffeting \citep{Kuniyoshi_2023_ApJ, Kuniyoshi_2025_ApJ} and are thus more prone to steepening into shocks via nonlinear mode conversion.
A more detailed investigation is left for future work.

\section{Discussions}\label{sec:discussion}

\subsection{Comparison with previous magneto-convection simulations}
\label{subsec:comparison_with_previous_magneto_convection_simulations}

Our simulation shows that the corona is heated in an AC-like manner.
This contrasts with several previous magneto-convection simulations that found the corona is heated in a DC-like manner \citep{Hansteen_2015_ApJ, Rempel_2017_ApJ}.
The discrepancy can be attributed to either or both of the following limitations of those studies.

First, their relatively short coronal loops \citep[$<50\ \rm Mm$,][]{Hansteen_2015_ApJ} 
have Alfv\'{e}n crossing times shorter than the convective timescale, 
and therefore tend to fall into the DC regime.
Second, their coarser spatial resolution \citep[$>100\ \rm km$,][]{Rempel_2017_ApJ} may have underestimated the Alfv{\'e}nic wave energy flux. A previous study showed that degrading resolution from $60\ \rm km$ to $200\ \rm km$ reduces the upward photospheric Poynting flux by a factor of about $10$ \citep{Yadav_2020_ApJ}.
Furthermore, insufficient resolution may suppress turbulent cascades, causing 
waves to be numerically dissipated before strongly contributing to coronal heating.

\subsection{Comparison with observations of coronal Alfv{\'e}nic waves}
\label{subsec:comparison_with_observations_of_coronal_alfvenic_waves}

Our simulation exhibits a few inconsistencies with observations that should be 
addressed in future work. First, the correlation length of Alfv{\'e}nic waves reaches up to $0.8\ \rm Mm$ 
(Figure~\ref{fig:msc_corr_length}(b)), $5$--$10$ smaller than recent DKIST\footnote{Daniel K. Inouye Telescope \citep{Rimmele_2020_SoPh}}/Cryo-NIRSP\footnote{Cryogenic Near-Infrared Spectropolarimeter \citep{Fehlmann_2023_SoPh}} measurements \citep{Morton_2025a_ApJ, Hahn_2025_ApJ}. This likely reflects the limited horizontal domain ($9 \times 9\ \rm Mm^2$), which suppresses supergranulation-scale organization of the surface magnetic field \citep[$15$--$30\ \rm Mm$,][]{BellotRubio_2019_LRSP}. 
Indeed, previous observations have suggested that the spatial coherence of Alfv{\'e}nic waves is associated with supergranulation \citep{Fargette_2021_ApJ, Sharma_2023_NatAs}.

Second, the PSDs of Alfv{\'e}nic waves in our simulation are not fully consistent with observations.
A recent DKIST/Cryo-NIRSP observation reports a skewed log-normal component in the coronal Alfv{\'e}nic wave PSD, peaking around $f = 3.5\ \rm mHz$ \citep{Morton_2025c_ApJ}, whereas our simulation peaks at a higher frequency 
of $f \approx 10\ \rm mHz$.
This discrepancy likely arises because partial ionization effects, which efficiently dissipate Alfv{\'e}nic waves above $f \approx 10\ \rm mHz$ \citep{Soler_2017_ApJ, Soler_2019_ApJ}, are not included in our simulation. Incorporating them following \citet{Khomenko_2018_AA} is left for future work.
In addition, our simulation does not fully resolve the transition region \citep[thickness $\lesssim 10\ \rm km$,][]{Bradshaw_2013_ApJ}, which modifies the transmission properties of Alfv{\'e}nic waves \citep{Howson_2023_MNRAS}. 
This may be mitigated using transition region broadening techniques \citep{Johnston_2019_ApJ, Johnston_2021_AA, Iijima_2021_ApJ}.

Our simulation shows that the coronal loop supports both propagating and standing Alfv{\'e}nic waves.
While propagating modes have been observed in quiet-Sun loops \citep{Tiwari_2021_ApJ, Morton_2021_ApJ}, observational evidence for standing modes in such loops is still lacking. 
DKIST/Cryo-NIRSP, with its high spatio-temporal resolution, may resolve this discrepancy by capturing the small-scale and high-frequency Alfv{\'e}nic waves associated with standing modes.

\subsection{Phenomenological heating rate from Alfv{\'e}n wave turbulence}
\label{subsec:phenomenological_heating_rate_from_alfven_wave_turbulence}

The phenomenological heating rate based on the Alfv{\'e}n wave turbulence model contains a free parameter, $c_{\rm d}$ (see Equation~\eqref{eq:awt_heating_rate}). 
The specific value of $c_{\rm d}$ remains under debate, with values reported in the range $0.1$--$1$ \citep[e.g.,][]{Hossain_1995_PhFl, vanBallegooijen_2011_ApJ, Chandran_2019_JPlPh, Verdini_2019_SoPh}.

Our inferred $c_{\rm d} \approx 1.3$ should be regarded as an upper limit, because we employ high-pass-filtered Els{\"a}sser variables to calculate $Q_{\rm awt}$ (Equation~\eqref{eq:awt_heating_rate}), which excludes low-frequency Alfv{\'e}nic fluctuations where AC and DC contributions coexist.
$c_{\rm d}$ is sensitive to the cutoff frequency of the high-pass filter, since the phenomenological heating rate $Q_{\rm awt}$ scales with the cube of the filtered Els{\"a}sser variables. 
Nevertheless our inferred values remain of order unity, consistent with the previous estimates.

\subsection{Numerical resolution}
\label{subsec:numerical_resolution}

Our simulation uses a grid spacing of $60 \ \rm km$, which does not resolve the physical dissipation scales in the corona \citep{Peter_2015_RSPTA}. In addition, we do not include explicit resistivity or viscosity in our simulation.
Nevertheless, the dominance of Alfv{\'e}n wave turbulence as the primary wave damping mechanism \added{is likely to hold} for the following reasons.

First, although heating is estimated via numerical dissipation, this remains physically meaningful because the dissipative structures originate from physical processes, i.e., energy cascade via Alfv{\'e}n wave turbulence. Furthermore, the energy cascade rate of Alfv{\'e}n wave turbulence is independent of the dissipation scale, so the predicted wave damping timescale does not drastically change with grid spacing.

Second, other damping mechanisms have timescales more than ten times longer than that of Alfv{\'e}n wave turbulence (Figure~\ref{fig:timescales}). Shortening these timescales would require smaller density clump radii, higher velocity amplitudes, or greater density contrast. However, the physical conditions in our simulation are already consistent with coronal loop observations (Section~\ref{subsec:density_inhomogeneity}), leaving little room for drastic changes. We therefore do not expect finer resolution to significantly alter this conclusion.

Third, increasing the resolution would more accurately resolve the transition region, producing a steeper Alfv{\'e}n speed gradient that causes upward-propagating waves to dissipate before reaching the corona, thereby further reducing the shock heating contribution relative to Alfv{\'e}n wave turbulence in the corona (Figure~\ref{fig:compressible_rate}).

% While the dominance of Alfv{\'e}n wave turbulence is robust, we acknowledge that the precise value of the contribution to the entire coronal heating ($\ge 80 \%$) may shift somewhat with resolution.  Therefore, magneto-convection simulations with finer grid spacing are required to determine how this fraction converges with resolution.

\added{While Alfvén wave turbulence is likely to remain the dominant damping mechanism, the precise value of its contribution to the entire coronal heating ($\ge 80 \%$) should be regarded as an estimate for the present simulation rather than a fully converged value. Higher-resolution simulations are required to determine how this fraction converges with resolution.}

\section{Conclusion}
\label{sec:conclusion}

The heating of coronal loops has been a long-standing open question. 
Many previous simulations have focused on active region loops, where the 
mean coronal magnetic field reaches $\sim60$--$100$~G, whereas the quiet 
Sun is also filled with coronal loops with much weaker fields 
($\lesssim10$~G).
The importance of developing self-consistent models of coronal heating has 
been emphasized in recent reviews \citep{VanDoorsselaere_2020_SSRv, 
Morton_2023_RvMPP}.

\medskip

Here we have developed a self-consistent numerical model of a quiet-Sun 
coronal loop driven by surface convection.
Our results demonstrate that the corona is heated predominantly through 
the dissipation of Alfv{\'e}nic waves, i.e., AC heating.
\added{In our simulation}, Alfv\'{e}n wave turbulence is the dominant wave dissipation mechanism, 
accounting for at least $80\%$ of the entire coronal heating and 
approximately an order of magnitude more effective than other mechanisms: 
phase mixing, resonant absorption, uniturbulence, KH instability, 
and Alfv\'{e}nic wave induced shock heating.
While prior studies have focused on Alfv\'en wave 
turbulence in isolation and shown it can heat the solar corona, 
the present work self-consistently 
demonstrates its dominance over other dissipation mechanisms 
under realistic coronal conditions including compressibility and density inhomogeneity.
Our results strongly support Alfv{\'e}n wave turbulence--based models used in 
space weather and stellar activity research, such as the Alfv{\'e}n Wave Solar 
Model \citep[AWSoM,][]{vanDerHolst_2014_ApJ} and the Magnetohydrodynamic 
Algorithm outside a Sphere \citep[MAS,][]{Mikic_2018_NatAs}.

\medskip 

This study provides a framework for analyzing self-consistent Alfv{\'e}nic 
wave heating in magneto-convection simulations, applied here to a single 
quiet-Sun coronal loop.
In reality, coronal loops exhibit a wide range of magnetic field strengths and 
lengths, and extending this analysis to diverse loop parameters and open field 
regions will be an important next step toward a comprehensive understanding of 
Alfv{\'e}nic wave heating in the solar corona.

%% Please use the acknowledgment and contribution environments. This will 
%% be anonomyized when the "anonymous" style option is used. 
\begin{acknowledgments}

The authors would like to thank Haruhisa Iijima for providing the RAMENS code.
Numerical simulations were carried out
using the Cray XD2000 system at the Center for Computational
Astrophysics (CfCA), National Astronomical Observatory of
Japan, and the A-KDK computer system at the Research
Institute for Sustainable Humanosphere, Kyoto University.
H.K. gratefully acknowledges support from the Japan Society for the Promotion of Science (JSPS) Overseas Research Fellowship. 
R.J.M. is supported by the UKRI Future Leaders Fellowship (RiPSAW MR/T019891/1 and MR/Z000289/1).

\end{acknowledgments}

\appendix

\section{Velocity power spectrum in the photosphere and corona}\label{sec:appendix}

\begin{figure}[t!]
  \centering
  \includegraphics[width=8cm]{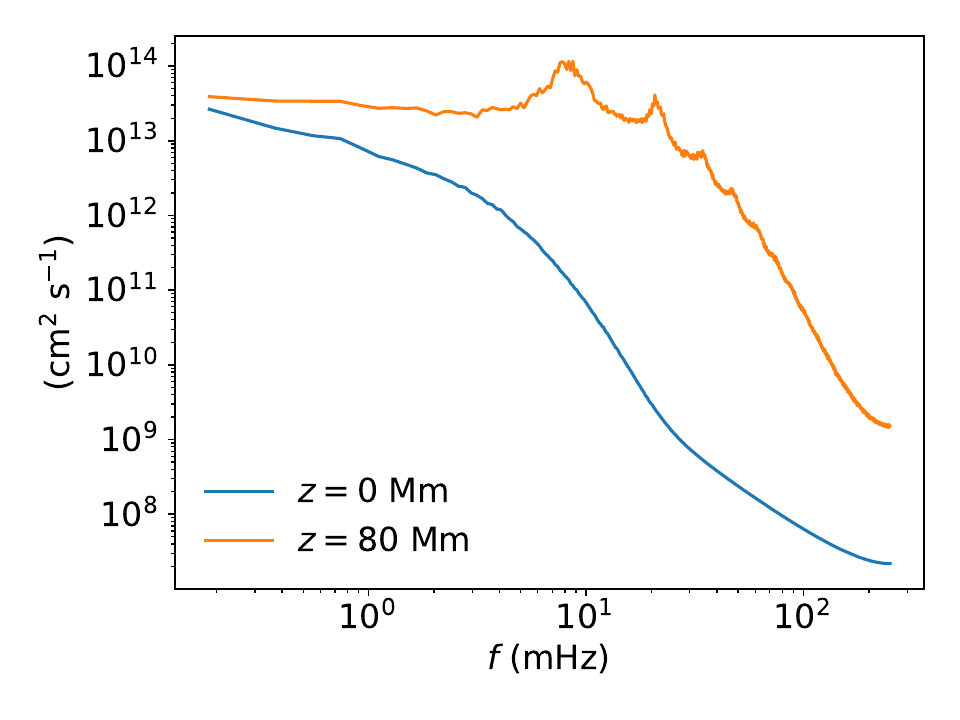} % 幅を10cmに指定
  \caption{
  The $xy$-averaged PSDs of the horizontal velocity ($\langle \hat{E}_{\perp}^{\rm k} \rangle_{xy}$) as a function of frequency $f$. 
  The blue line shows the PSD at $z=0\ \mathrm{Mm}$ in the photosphere. 
  The orange line shows the PSD at $z=80\ \mathrm{Mm}$ in the corona.
  }
  \label{fig:ph_psd}
\end{figure}

\added{Figure~\ref{fig:ph_psd} shows the $xy$-averaged PSD of the horizontal velocity as a function of frequency $f$ ($\langle \hat{E}_{\perp}^{\rm k} \rangle_{xy}$) in the photosphere.
Here, the spectrum at $z=0\ \mathrm{Mm}$ is shown as a representative photospheric example; we confirmed that the photospheric layer on the other side of the domain, $z=160\ \mathrm{Mm}$, exhibits a similar PSD profile.
We emphasize that this photospheric PSD is not imposed as a driver, but emerges self-consistently from the convective motions.
For reference, Figure~\ref{fig:ph_psd} also includes the PSD at $z=80\ \mathrm{Mm}$ in the corona.}

\added{The photospheric and coronal spectra differ as a result of several processes acting on the waves as they propagate upward: 
amplification of the wave amplitudes with height due to density stratification, 
reflection in the chromosphere and transition region, which preferentially affects the lower-frequency components \citep{Cranmer_2005_ApJS, Soler_2017_ApJ}, 
transport of energy from lower to higher frequencies through the turbulent energy cascade, 
dissipation of wave energy, 
and resonance in the corona.}

%% For this sample we use BibTeX plus aasjournalv7.bst to generate the

% \bibliography{sample701}{}
% \bibliographystyle{aasjournalv7}
% ------------------------

% ------------------------


% Include this line if you are using the \added, \replaced, \deleted
% commands to see a summary list of all changes at the end of the article.
% \listofchanges

\end{document}